\pdfoutput=1
\documentclass[11pt]{article}
\usepackage{jheppub}

\newcommand{\be}{\begin{equation}}
\newcommand{\ee}{\end{equation}}
\newcommand{\bea}{\begin{eqnarray}}
\newcommand{\eea}{\end{eqnarray}}
\def\eqn#1{eq.~(\ref{#1})}

\def\eqns#1#2{eqs.~(\ref{#1}) and~(\ref{#2})}

\newcommand{\Li}{\textrm{Li}}
\newcommand{\Gcusp}{\Gamma_{\rm cusp}}

\newcommand{\ket}[1]{\langle #1 \rangle}

\newcommand{\cA}{\mathcal{A}}

\newcommand{\cC}{\mathcal{C}}
\newcommand{\cE}{\mathcal{E}}
\newcommand{\cF}{\mathcal{F}}
\newcommand{\cG}{\mathcal{G}}

\newcommand{\cN}{\mathcal{N}}
\newcommand{\cO}{\mathcal{O}}
\newcommand{\cP}{\mathcal{P}}
\newcommand{\cR}{\mathcal{R}}
\newcommand{\cS}{\mathcal{S}}
\newcommand{\cZ}{\mathcal{Z}}

\DeclareMathOperator*{\res}{Res}
\newcommand{\CO}[2]{CO$^{(u_#1,u_#2)}$}

\title{Heptagon Functions and Seven-Gluon Amplitudes in Multi-Regge Kinematics}

\author{Lance~J.~Dixon,$^{1}$ }
\author{Yu-Ting~Liu$^{1}$ and }
\author{Julian~Miczajka$^{2,3}$ }

\affiliation{$^1$ SLAC National Accelerator Laboratory,
Stanford University, Stanford, CA 94309, USA}
\affiliation{$^2$ Institut f\"ur Physik, Humboldt-Universit\"at zu Berlin,
Zum Gro\ss en Windkanal 6, D-12489 Berlin, Germany}
\affiliation{$^3$ Max-Planck-Institut f\"ur Physik, F\"ohringer Ring 6, D-80805 M\"unchen, Germany}

\abstract{We compute all $2\to5$ gluon scattering amplitudes in planar $\cN=4$ super-Yang-Mills theory in the multi-Regge limit that is sensitive to the non-trivial (``long'') Regge cut. We provide the amplitudes through four loops and to all logarithmic accuracy at leading power, in terms of single-valued multiple polylogarithms of two variables.  To obtain these results, we leverage the function-level results for the amplitudes in the Steinmann cluster bootstrap. To high powers in the series expansion in the two variables, our results agree with the recently conjectured all-order \emph{central emission vertex} used in the Fourier-Mellin representation of amplitudes in multi-Regge kinematics. Our results therefore provide a resummation of the Fourier-Mellin residues into single-valued polylogarithms, and constitute an important cross-check between the bootstrap approach and the all-orders multi-Regge proposal.
}

\emailAdd{lance@slac.stanford.edu}
\emailAdd{aytliu@stanford.edu}
\emailAdd{miczajka@mpp.mpg.de}

\preprint{ \begin{flushright} SLAC--PUB--17630 \\ HU-EP-21/43 \end{flushright}}

\begin{document}
\maketitle
\flushbottom


\section{Introduction}
The study of Regge limits in relativistic particle scattering can be traced back to the early analytic $S$-matrix program (see e.g.~Ref.~\cite{ELOP}), in which general properties of scattering are analyzed using only first principles such as unitarity, analyticity, and crossing symmetry.  For example, bounds can be placed on total cross sections at high energy through the optical theorem. Another example is the existence of Regge poles, where a series of bound states in one scattering channel can be related to high-energy behavior in a different channel. The Regge limit has also been studied extensively in the context of perturbative quantum field theories. One such example is the BFKL equation in QCD \cite{Fadin:1975cb,Kuraev:1976ge,Lipatov:1976zz,Kuraev:1977fs,Balitsky:1978ic}, which resums logarithms of large rapidity intervals in parton-parton scattering (see also Refs.~\cite{Grisaru:1973vw,Grisaru:1973wbb}). Other elements of the BFKL approach include impact factors and central emission (Lipatov) vertices, for coupling to the Reggeized gluon ladder at one end or in the middle.  Often these elements can be extracted from fixed order amplitudes. For example, the one-loop central emission vertex in QCD can be extracted from the one-loop five-gluon amplitude~\cite{DelDuca:1998cx}. The main focus of this paper will be multi-Regge limits in planar $\cN=4$ super-Yang-Mills (SYM) theory. We will perform a study, analogous to Ref.~\cite{DelDuca:1998cx}, of a proposal for the all-orders central emission vertex in this theory by using the seven-point amplitude.

Planar $\cN=4$ SYM finds its stage in a modern revival of the analytic approach to scattering amplitudes, with which amplitudes can be calculated to unprecedentedly high loop orders in perturbation theory. (See Ref.~\cite{Caron-Huot:2020bkp} for a review.) In such an analytic \emph{bootstrap} program, one starts with a knowledge (or suspicion) about the space of functions to which the amplitudes belong. For example, maximally helicity violating (MHV) and next-to-maximally helicity violating (NMHV) amplitudes in planar $\cN=4$ SYM are expected to be multiple polylogarithms~\cite{Goncharov:2001iea} of weight $2L$ at $L$ loops~\cite{Arkani-Hamed:2012zlh}. Furthermore, their \emph{symbol letters}~\cite{Goncharov:2010jf} are conjectured to belong to the $\cA$-coordinates of certain cluster algebras~\cite{Golden:2013xva}. Armed with such knowledge, one can write down the most general ansatz and then strive to fix all unknown coefficients with known analytic and physical behaviors, such as (consecutive) branch cuts and soft, (multi-)collinear, and/or multi-Regge kinematic limits.

Through seven external particles, there are only MHV and NMHV helicity configurations (or their parity conjugates, $\overline{\text{MHV}}$ and $\overline{\text{NMHV}}$).  The \emph{symbol} of these amplitudes has been known through four loops for a few years by utilizing (at higher loops) the Steinmann cluster bootstrap~\cite{Caron-Huot:2011zgw,Drummond:2014ffa,Dixon:2016nkn,Drummond:2018caf,Caron-Huot:2020bkp}.  Recently, the \emph{beyond-the-symbol} information was filled in as well~\cite{Dixon:2020cnr}.  As a result, all iterated derivatives of the amplitudes are known, as well as boundary conditions for integrating up the derivatives.  In principle, the amplitudes can now be obtained for general kinematics, either analytically or numerically, by integrating from a known boundary point. In this heptagon bootstrap, no information from multi-Regge kinematics (MRK) was required.  Therefore the MRK limit is predicted unambiguously by the bootstrapped amplitudes.

The behavior of amplitudes in the MRK limit has also been explored using the BFKL approach.  In the unphysical Euclidean region, the MRK limit is equivalent to a soft limit, and (suitably-normalized) amplitudes behave smoothly there; i.e.~the remainder function vanishes.  In contrast, on physical scattering sheets, starting with six-particle scattering, there are non-trivial Regge cuts that dictate nonvanishing discontinuities for the amplitudes~\cite{Bartels:2008ce}. However, the cuts themselves satisfy BFKL evolution equations and factorize after performing a Fourier-Mellin transform~\cite{Bartels:2009vkz,Lipatov:2009nt,Lipatov:2010qf,Lipatov:2010qg,Lipatov:2010ad,Bartels:2011ge,Fadin:2011we,Dixon:2012yy,DelDuca:2016lad,DelDuca:2018hrv,DelDuca:2019tur,Bartels:2020twc}.  The amplitude in momentum space is given by the inverse Fourier-Mellin transform, which includes a Fourier sum and an integral over the Mellin moment (two of each in the seven-point case).  The integrand of the inverse Fourier-Mellin sum-integral contains three building blocks:  Two of them, the BFKL eigenvalue and the impact factor, are known to all orders in the 't Hooft coupling through a duality between scattering amplitudes and Wilson loops in $\cN=4$ SYM and the all-orders flux-tube formalism~\cite{Basso:2014pla}. These two components are sufficient for determining all six-point amplitudes in MRK. They have been tested against amplitudes from the hexagon function bootstrap through seven loops~\cite{Caron-Huot:2019vjl}.  At seven points, a third component is required, the \emph{central emission vertex}; its value to all orders was recently conjectured~\cite{DelDuca:2019tur}.

In this paper, we compute through four loops all $2\to5$ gluon scattering amplitudes in MRK with the non-trivial (``long'') Regge cut~\cite{Bartels:2011ge,Bartels:2013jna,Bartels:2014jya}, by integrating along a sequence of paths connecting a known boundary point to MRK, using the iterated derivatives and boundary conditions obtained in Ref.~\cite{Dixon:2020cnr}. We choose this ``long and winding road'' so that the functions we need to integrate along each path in the sequence (until the end) are relatively simple, usually harmonic polylogarithms (HPLs)~\cite{Remiddi:1999ew} or $G$-functions~\cite{Goncharov:1998kja} of a single argument, with indices 0 or 1 only.  The final MRK results at each loop order can be expressed as a set of coefficient functions multiplying two large logarithms; the coefficient functions are single-valued multiple polylogarithms of definite weight in two complex variables, $\rho_1$ and $\rho_2$.  We then compare our results to the Fourier-Mellin sum-integrals incorporating the conjectured central emission vertex~\cite{DelDuca:2019tur}.  The integrals over the Mellin moments can be converted to sums by deforming the integrals and extracting residues~\cite{DelDuca:2018hrv}.  Rather than performing the multiple sums directly, we series expand the multiple polylogarithms to high powers in $\rho_1$ and $1/\rho_2$, and we find complete agreement with the conjecture.

The paper is organized as follows. In Section~\ref{sec:review} we review the general structure of amplitudes in planar $\cN=4$ SYM.  We describe how the bootstrapped amplitudes are represented iteratively in terms of heptagon functions, using a Hopf algebra coproduct, and how one can obtain the amplitudes in general kinematics through integration.  We also review multi-Regge kinematics as well as another special kinematics, the \emph{collinear-origin} (CO) surface, on which the boundary conditions are specified for the bootstrapped amplitudes~\cite{Dixon:2020cnr}. In Section~\ref{sec:longroad}, we describe a sequence of paths that connects the boundary CO surface (which is on the Euclidean Riemann sheet) to MRK (on the physical sheet for $2\to5$ scattering). We integrate along these paths starting with the CO surface boundary conditions, and eventually we obtain the amplitudes in MRK. In each intermediate step the amplitudes are obtained in closed form; we also discuss the function space the amplitudes belong to on these paths and in MRK.

In Section~\ref{sec:FM} we examine the inverse Fourier-Mellin sum-integral used in the literature to compute amplitudes directly in MRK. The non-trivial part of the integral can be represented as a sum of multiple residues, which in turn gives a series expansion in two complex variables, $\rho_1$ and $1/\rho_2$. We compare the series expansion against our results from the bootstrapped amplitudes, and obtain agreement to high powers. Finally, in Section~\ref{sec:line} we examine the behavior of the MRK coefficient functions on a particular line through the four-dimensional space $(\rho_1,\rho_2,\bar\rho_1,\bar\rho_2)$, where $\rho_1=\bar\rho_1=1/\rho_2=1/\bar\rho_2$.  This line preserves both parity and a target-projectile discrete symmetry.  The coefficient functions on this line collapse to HPLs in a single variable with weights $\{0,1,-1\}$, which we provide explicitly and also plot.

This paper also contains five appendices.  Appendices~\ref{app:bds} and~\ref{app:bdslike} define the BDS and BDS-like amplitude normalizations.
Appendix~\ref{app:MRKsym} describes the action of parity and target-projectile
discrete symmetries on the MRK variables.  Appendix~\ref{app:polylog} contains
a short primer on (single-valued) multiple polylogarithms.  Finally,
appendix~\ref{sec:letterlimits} records how the 42 symbol letters for heptagon
functions behave in the MRK limit, as well as on the CO surface and on the
other intermediate path segments between these two regions.  This information is necessary for integrating up the functions from their coproducts on each segment.

We also provide a set of computer-readable ancillary files.
The file {\tt g\_h\_MHV.m} provides the coefficient functions for the MHV
amplitude $({+}{+}{+})$ in MRK in terms of single-valued $\cG$-functions, while
{\tt g\_h\_NMHV\_mpp.m} and {\tt g\_h\_NMHV\_pmp.m} do the same for
the two independent NMHV helicity configurations,
$({-}{+}{+})$ and $({+}{-}{+})$. Furthermore, {\tt g\_h\_TPS.m} contains the expressions for all MHV and NMHV coefficient functions on the target-projectile-parity symmetric line in terms of harmonic polylogarithms. For the readers' convenience, we also provide a file {\tt cGToG.m} that converts single-valued $\cG$-functions to combinations of $G$-functions, in a basis suitable for series expansion.


\section{Amplitudes and Kinematics} \label{sec:review}

\subsection{General kinematics and normalization}
\label{sec:kinnorm}

In planar $\cN=4$ SYM, we work with color-ordered partial amplitudes, which are coefficients of single traces $\text{Tr}(T^{a_1}T^{a_2}\cdots T^{a_n})$ in the color decomposition. Furthermore, we package amplitudes related by supersymmetric Ward identities into a \emph{superamplitude} $\cA_n(\Phi_i,\ i=1\ldots n)$, with on-shell superfields $\Phi_i = \Phi(k_i,\eta_i)$ where
\be
\Phi(k,\eta) = G^+ +\eta^A\Gamma_A+\tfrac{1}{2!}\eta^A\eta^B S_{AB}+\tfrac{1}{3!}\eta^A\eta^B\eta^C\epsilon_{ABCD}\bar \Gamma^D+\tfrac{1}{4!}\eta^A\eta^B\eta^C\eta^D\epsilon_{ABCD}G^- .
\label{eq:superfield}
\ee
Here $G^\pm$ are gluon fields of plus/minus helicity, $\Gamma_A,\bar\Gamma^D$ the gluinos and $S_{AB}$ the scalars.
The expansion of $\cA_n$ in Grassmann variables $\eta_i$ naturally organizes it into components in an N$^k$MHV expansion,
\be\label{eq:NkMHVexpansion}
\cA_n = \cA_n^{\text{MHV}} + \cA_n^{\text{NMHV}} + \cA_n^{\text{NNMHV}} + \dots + \cA_n^{\overline{\text{MHV}}} \,,
\ee
where each successive term contains four more $\eta$s.
For seven particles, the full superamplitude $\cA_7$ is determined by $\cA_7^{\text{MHV}}$ and $\cA_7^{\text{NMHV}}$, because the remaining two components, $\cA_n^{\overline{\text{NMHV}}}$ and $\cA_n^{\overline{\text{MHV}}}$, are fixed by parity conjugation. To obtain gluonic amplitudes, one simply needs to perform relevant projections with respect to the Grassmann variables $\eta_i$; a negative helicity gluon requires $(\eta_i)^4$ according to \eqn{eq:superfield}.

The full amplitude $\cA_n$ factors into an infrared-divergent part, the BDS ansatz $\cA_n^{\text{BDS}}$ (see Appendix~\ref{app:bds}), and an infrared-finite part $\cR_n$,
\be
\cA_n = \cA_n^{\text{BDS}} \cdot \cR_n \,.
\ee
The infrared-finite $\cR_n$ will be referred to as the \emph{BDS-normalized amplitude}. It has the same N$^k$MHV decomposition as the full amplitude, $\cR_n = \cR_n^{\text{MHV}} + \cR_n^{\text{NMHV}} + \dots$, and it respects (dual) super-conformal symmetries. As a result, its bosonic part only depends on conformal cross ratios,
\be
u_{i,j} = \frac{x_{i,j+1}^2\ x_{i+1,j}^2}{x_{i,j}^2\ x_{i+1,j+1}^2}\,,
\ee
where $x_{i,j} \equiv x_i - x_j$ are dual coordinates defined by $p_i = x_{i+1}-x_i$, and $p_i$ is the (lightlike) momentum of the $i^{\rm th}$ particle; in addition, momentum conservation requires $x_{i+n} \equiv x_i$.
For $n=7$, we have seven cross ratios,
\be
u_i \equiv u_{i+1,i+4} \,,\quad 1\le i\le 7 \,,
\ee
which satisfy one Gram determinant constraint,
\bea
\label{eq:gramdet}
0 = 1 + \biggl[-u_1 &+& u_1 u_3 + u_1 u_4 + u_1 u_2 u_5 - u_1 u_3 u_5 - u_1^2 u_4 u_5 - 2 u_1 u_2 u_4 u_5 \nonumber\\
  &+& u_1 u_2 u_3 u_5 u_6 + u_1^2 u_2 u_4 u_5^2 \ +\ \text{cyclic}\biggr] + u_1 u_2 u_3 u_4 u_5 u_6 u_7 \,,
\eea
and therefore only six cross ratios are independent. In terms of Mandelstam invariants $s_{i,i+1} = (p_i+p_{i+1})^2$ and $s_{i,i+1,i+2} = (p_i+p_{i+1}+p_{i+2})^2$, the cross ratios are
\bea
u_1 &=& \frac{s_{34}s_{671}}{s_{234}s_{345}} \,, \qquad
u_2 = \frac{s_{45}s_{712}}{s_{345}s_{456}} \,, \qquad
u_3 = \frac{s_{56}s_{123}}{s_{456}s_{567}} \,, \qquad
u_4 = \frac{s_{67}s_{234}}{s_{567}s_{671}} \,, \nonumber\\[7pt]
u_5 &=& \frac{s_{71}s_{345}}{s_{671}s_{712}} \,, \qquad
u_6 = \frac{s_{12}s_{456}}{s_{712}s_{123}} \,, \qquad
u_7 = \frac{s_{23}s_{567}}{s_{123}s_{234}} \,. 
\label{eq:uisi}
\eea

The kinematics can also be described using \emph{momentum (super)twistors}~\cite{Hodges:2009hk,Mason:2009qx},
\be
\cZ_i = (Z_i \, | \, \chi_i),
\ee
where $Z_i\in \mathbb{P}^3$ are bosonic momentum twistors and $\chi_i$ their fermionic counterparts. Bosonic invariants are constructed from momentum twistor four-brackets,
$\left<ijkl\right> \equiv \det(Z_i Z_j Z_k Z_l)$.  A subset of the four-brackets are related to the dual coordinates,
\be
x_{i,j}^2
= \frac{\left<i-1,i,j-1,j\right>}{\left<i-1,i\right> \left<j-1,j\right>} \,,
\ee
where $\left<ij\right> = \det(\lambda_i\lambda_j)$ is the usual spinor product.
The Grassmann information in the superamplitude is encoded in the $R$-invariants, or five-brackets, defined as
\be\label{eq:5bra}
[abcde] = \frac{\delta^{0|4} \big( \chi_a \langle b c d e \rangle + \text{cyclic} \big)}{\langle a b c d \rangle \langle b c d e \rangle \langle c d e a \rangle \langle d e a b \rangle \langle e a b c \rangle} \,,
\ee
where `$+$~cyclic' means summing over all 5 cyclic rotations generated by
$a\to b\to c\to d\to e\to a$.

For seven particles, we adopt the notation of Ref.~\cite{Dixon:2016nkn} and write the five-bracket in terms of the two omitted labels,
\be
(67) = (76) \equiv [12345], \quad\quad \text{etc.}
\ee
The MHV amplitude $\cR_7^{\text{MHV}}$ is free of Grassmann variables and is therefore a single bosonic function of the cross ratios.  It is invariant under the dihedral symmetry group $D_7$, which includes cyclic transformations of order 7, and a reflection or flip.  (In the notation of Ref.~\cite{Dixon:2016nkn}, $\cR_7^{\text{MHV}} = \mathcal{B}_7$.)  On the other hand, the NMHV amplitude $\cR_7^{\text{NMHV}}$ is a linear combination of five-brackets~(\ref{eq:5bra}), with conformally invariant bosonic coefficients.  The $\binom{7}{2}=21$ five-brackets are not all independent but satisfy 6 linear relations.  The $15 = 2 \times 7 + 1$ independent ones can be chosen to be $(12)$, $(14)$, their cyclic images, and the dihedrally-invariant tree amplitude, leading to the decomposition
\be \label{eq:NMHVcomponents}
\cR_7^{\text{NMHV}} =
\left[\frac{3}{7}(12) + \frac{1}{7}(13) + \frac{2}{7}(14) + \text{cyclic}\right] B_0 +
\bigg[(12) B_{12} + (14) B_{14} + \text{cyclic} \bigg] \,,
\ee
where $B_0,B_{12},\ldots$ are referred to as ``components'' of the BDS-normalized NMHV amplitude~\cite{Dixon:2016nkn}. In the rest of the paper, we restrict our discussion to seven particles and will often drop the subscript $n=7$.


\subsection{Amplitudes from coproducts}

The bosonic functions in the BDS-normalized amplitude $\cR$ are multiple polylogarithms of weight $2L$ at $L$ loops. In practice we first calculate the BDS-like normalized amplitudes~(\ref{eq:bdslikeampmhv}) and~(\ref{eq:bdslikeampnmhv}) in a given kinematical region, because they belong to a smaller function space (see Appendix~\ref{app:bdslike}). Then we convert them to $\cR$ using \eqns{eq:bdslikeampmhv}{eq:bdslikeampnmhv}, and the known expressions for $\Gcusp$ and $\cE^{(1)}$ in that region. The seven-point amplitudes through four loops are represented in Ref.~\cite{Dixon:2020cnr} as iterated coproducts, which describe their first derivatives iteratively in any kinematical region. In addition, the boundary conditions are provided for all functions on a particular surface, the Euclidean \emph{collinear-origin} surface (see Section~\ref{sec:CO}).

The heptagon symbol alphabet consists of the following 42 $g$-letters\footnote{The Pl\"ucker bilinear is defined as $\ket{a(bc)(de)(fg)} \equiv \ket{abde}\ket{acfg}-\ket{abfg}\ket{acde}$.}, 
\begin{align}
\label{eq:gletters}
g_{11} &= \frac{\ket{1256}\ket{2345}}{\ket{1245}\ket{2356}}
 \,=\, u_1 \,, \quad & \quad
g_{41} &= \frac{\ket{1346}\ket{7(12)(34)(56)}}{\ket{1256}\ket{1347}\ket{3467}}\,, \nonumber \\[7pt]
g_{21} &= \frac{\ket{1235}\ket{2456}}{\ket{1245}\ket{2356}}
  \,=\, 1-u_1 \,, \quad & \quad
g_{51} &= \frac{\ket{1235}\ket{1267}\ket{4567}}{\ket{1237}\ket{1567}\ket{2456}}\,, \\[7pt]
g_{31} &= \frac{\ket{7(23)(45)(61)}}{\ket{1457}\ket{2367}}
  \,=\, 1-u_3 u_6 \,, \quad & \quad
g_{61} &= \frac{\ket{1237}\ket{1346}\ket{2345}\ket{4567}}{\ket{1234}\ket{3456}\ket{7(12)(34)(56)}}\,, \nonumber
\end{align}
plus their cyclic permutations $g_{ij} \equiv g_{i1}\big |_{Z_{k} \to Z_{k+j-1}}$, $j=1,2,\ldots,7$. (The letters $g_{4j},g_{5j},g_{6j},$ have more complicated representations in terms of the cross ratios $u_i$, which we omit here.)

Coproducts encode derivatives. To be specific, for a function $F$ let us use $F^a$ to denote its coproduct with respect to letter $a$. Then the derivative of $F$ with respect to a kinematic variable $x$ is given by
\be
  \frac{\partial F}{\partial x} = \sum_{a\in\cS} F^a \left(\frac{\partial}{\partial x} \ln a \right) \,,
\ee
where the sum is over the whole symbol alphabet $\cS$. The derivatives as specified above, together with the boundary conditions, can be used to reconstruct $F$ along any path from the boundary through integration. In general this can be done numerically. For special paths, the alphabet~(\ref{eq:gletters}) collapses to a simpler set of letters. Then the function $F$ can be reconstructed in closed form, for example, in terms of $G$-functions (see Appendix~\ref{app:polylog}) with simple arguments.


\subsection{Multi-Regge kinematics}

We will investigate $2\to5$ gluon scattering with particles 2 and 3 incoming, and particles 1 and 4 the two highest-energy outgoing gluons. In the center-of-mass frame, we have $p_2^+ = p_3^- = \mathbf{p}_2 = \mathbf{p}_3 = 0$ and $p_2^- = p_3^+ = \sqrt{s}$, where $p_i^\pm$ are light-cone coordinates and $\mathbf{p}_i$ the transverse momenta, defined via
\be
p^{\pm} = p^0 \pm p^3, \qquad \mathbf{p} = p^1 + i p^2.
\ee
In multi-Regge kinematics (MRK), the produced gluons are strongly ordered in rapidity with comparable transverse momenta,
\be
 p_4^+ \gg p_5^+ \gg p_6^+ \gg p_7^+ \gg p_1^+,
\ee
and
\be
|\mathbf{p_4}| \simeq |\mathbf{p_5}| \simeq |\mathbf{p_6}| \simeq |\mathbf{p_7}| \simeq |\mathbf{p_1}|\,.
\ee
Inserting this behavior into \eqn{eq:uisi}, we see that in MRK the cross ratios behave as
\bea
u_1\,,u_2\,,u_5\,,u_6 &\sim& \cO(\delta), \nonumber\\
1-u_3\,,1-u_4 &\sim& \cO(\delta), \nonumber\\
1-u_7 &\sim& \cO(\delta^2) \,,
\label{eq:MRKscaling}
\eea
where $\delta$ is some small parameter characterizing the infinitesimal ratios $p_{i+1}^+/p_i^+$ for $i=4,5,6,7$.

In the literature, the MRK limit is often parametrized by two small real parameters $\tau_1,\tau_2$ and two complex variables $z_1,z_2$.
In our conventions, these variables are related to the cross ratios by
\bea
\label{eq:MRKvariables}
\sqrt{u_1 u_2} &=& \tau_1 \,,  \qquad
\frac{u_1}{1-u_3} = \left|\frac{1}{1-z_1}\right|^2, \qquad
\frac{u_2}{1-u_3} = \left|\frac{z_1}{1-z_1}\right|^2,
\nonumber \\
\sqrt{u_5 u_6} &=& \tau_2 \,,  \qquad
\frac{u_5}{1-u_4} = \left|\frac{1}{1-z_2}\right|^2, \qquad
\frac{u_6}{1-u_4} = \left|\frac{z_2}{1-z_2}\right|^2 .
\eea
The \emph{simplicial coordinates} $\rho_1,\rho_2$ are also used \cite{Brown:2009qja,DelDuca:2016lad}; they are related to $z_1,z_2$ by
\be
\label{eq:simplicial}
z_1 = \frac{\rho_1 (1-\rho_2)}{\rho_1 - \rho_2}  \qquad
z_2 = \frac{\rho_2 - \rho_1}{1 - \rho_1} \,.
\ee
The limiting behavior of the entire heptagon alphabet in MRK is given in (\ref{eq:gsMRK}) and (\ref{eq:gsMRK2}).

In this paper, we look at the Mandelstam region where all the centrally-produced gluons (i.e.~particles 5, 6 and 7) are analytically continued to negative energies.  Equivalently, we take particles 1 and 4 to be incoming, and particles 2,3,5,6,7 to be outgoing.  From the Regge analysis, this is the region where the ``long'' Regge cut contributes~\cite{Bartels:2011ge,Bartels:2013jna,Bartels:2014jya}.  Other Mandelstam regions are not sensitive to this cut, but only to shorter cuts that are equivalent to cuts for the six-point amplitude, which do not involve the central emission vertex.  The long-cut region is related to the Euclidean region by analytically continuing $u_7$ once around its complex origin,
\be
\label{eq:continuation}
u_7 \to e^{-2\pi i}\,u_7 \,.
\ee

Different helicity configurations are denoted by $\cR_{h_1,h_2,h_3}$, where $h_1,h_2,h_3$ are the helicities of the centrally produced gluons (particles 5,6,7). There are only three independent amplitudes, $\cR_{+++}, \cR_{-++}$, and $\cR_{+-+}$, while all others are related to these by symmetry transformations (see Appendix~\ref{app:MRKsym}). $\cR_{+++}$ is just the MHV BDS-normalized amplitude; $\cR_{-++}$ and $\cR_{+-+}$ can be obtained from the NMHV super-amplitude by projecting on its Grassmann variables. We refer the reader to Refs.~\cite{DelDuca:2016lad,DelDuca:2018hrv} for the details of such projections; here we will simply quote the result for what \eqn{eq:NMHVcomponents} becomes for these components in MRK (after converting to our conventions):
\bea
\cR_{-++} &=& \big(B_0 + B_{71} + B_{12}\big) + R_{234}\big(B_{62}-B_{12}\big) + R_{235}\big(B_{67}-B_{62}\big) \,, \nonumber\\[10pt]
\cR_{+-+} &=& \big(B_0 + B_{14} + B_{25} + B_{51}\big)
  + R_{345}\big(B_{47}-B_{25}\big) \nonumber\\[5pt]
  &+& \overline{R}_{234}\big(B_{23}-B_{25}-B_{51}\big) + \overline{R}_{234} R_{345}\big(B_{73}-B_{23}+B_{25}\big) \,.
\eea
where $B_0,B_{12},\ldots$ are components in \eqn{eq:NMHVcomponents} and $R_{abc},\overline{R}_{abc}$ are given by 
\be
R_{234} = \frac{\rho_1(1-\rho_2)}{\rho_2(1-\rho_1)} \,,\quad
R_{235} = \frac{\rho_1}{\rho_1-1} \,,\quad
R_{345} = \frac{\rho_1-\rho_2}{1-\rho_2} \,,
\ee
and their complex conjugates.  In comparison with eq.~(5.62) of Ref.~\cite{DelDuca:2016lad}, their $\hat{X} \to B_0$ and their $\hat{V}_{ij}\to B_{i-1,j-1}$ due to different component indexing conventions; however, the $R_{abc}$ indexing, and the relation to the $\rho_{1,2}$ variables is identical.

In quoting the amplitudes, it is customary to factor out a pure imaginary phase from the BDS ansatz,
\be \label{eq:delta7}
\delta_7\ =\
\frac{\pi \Gcusp}{2} \ln \left[ \frac{\sqrt{u_1 u_2 u_5 u_6}}{1-u_7} \right]
\ =\
\frac{\pi \Gcusp}{4} \ln \left|\frac{\rho_1}{(1-\rho_1)(1-\rho_2)}\right|^2 \,,
\ee
where $\Gcusp$ is given in \eqn{eq:Gcusp}. The amplitudes are then doubly expanded in the coupling $g^2 = {g_{\text{YM}}^{2}N_c}/{(16 \pi^{2})}$ and in powers of large logarithms $\ln\tau_1, \ln\tau_2$ as follows,
\begin{multline}
\label{eq:ComponentExpansion}
\cR_{h_1,h_2,h_3}\, e^{i\delta_7} = 1 + 2\pi i \cdot
\sum_{L=1}^\infty g^{2L}
\sum_{n_1,n_2=0}^{n_1+n_2=L-1}
\left(\frac{1}{n_1!}\ln^{n_1} \tau_1\right)
\left(\frac{1}{n_2!}\ln^{n_2} \tau_2\right) \times \\[7pt]
\bigg[ g^{L;\,n_1,n_2}_{h_1,h_2,h_3}(\rho_1,\rho_2) +
2\pi i \cdot h^{L;\,n_1,n_2}_{h_1,h_2,h_3}(\rho_1,\rho_2) \bigg] \,,
\end{multline}
where the dependence on $g^2,\tau_1,\tau_2$ is explicit, and the dependence on $\rho_1,\rho_2$ is carried by $g^{L;\,n_1,n_2}_{h_1,h_2,h_3}$ and $h^{L;\,n_1,n_2}_{h_1,h_2,h_3}$.
The terms with $n_1+n_2=L-1$ contribute at the \emph{leading logarithmic accuracy} (LLA), those with $n_1+n_2=L-2$ contribute at next-to-LLA (NLLA), then N$^2$LLA, and so on. Since the $L$-loop amplitude as well as the $L$-loop contribution of the imaginary phase $\delta_7$ have uniform transcendentality $2L$, the component functions $g^{L;n_1,n_2}_{h_1,h_2,h_3}$ and $h^{L;n_1,n_2}_{h_1,h_2,h_3}$ are functions of uniform transcendental weight $2L-n_1-n_2-1$ and $2L - n_1 - n_2 -2$ respectively.


\subsection{Collinear-origin surface}
\label{sec:CO}
Here we review the collinear-origin (CO) kinematics. Amplitudes on the \emph{Euclidean} CO surface are known in closed form \cite{Dixon:2020cnr} and will serve as boundary conditions. The CO surface is described by the following kinematics,
\begin{align}
\label{eq:CO}
\textbf{CO$^{(u_{i+3},u_{i+4})}$} \quad
\begin{cases}
& u_{i+1},u_{i+2},u_{i+5},u_{i+6} \ll 1, \\
& u_{i+3},u_{i+4} \quad \text{generic}, \\
& 1-u_{i+7} = u_{i+1}(1-u_{i+4}) + u_{i+6}(1-u_{i+3}) \ll 1 \,.
\end{cases}
\end{align}
where $u_{i+7} \equiv u_{i}$ and the superscript $(u_{i+3},u_{i+4})$ denotes the two cross ratios which have generic, order-one, values on the surface. The final relation in \eqn{eq:CO} comes from expanding the Gram determinant relation~(\ref{eq:gramdet}) in the limit that $u_{i+1},u_{i+2},u_{i+5},u_{i+6}$ and $1-u_{i+7}$ are small.  There are 14 such CO surfaces, corresponding to letting $i=0,1,2,\ldots,6$, plus their parity-conjugated images which have the same values for the $u_i$.

On the \emph{physical} Riemann sheet, the kinematics \CO{3}{4} touches the MRK surface. Indeed, we can see that it satisfies \eqn{eq:MRKscaling} if we take $u_{3,4}\to1$. To be specific, the limit $u_{3,4}\to1$ on \CO{3}{4} matches MRK in the special limit $\rho_1\to0, \rho_2\to\infty, \bar{\rho}_1\to\bar{\rho}_2$ with the following identifications,
\bea \label{eq:MRKCOmatch}
&&\tau_1 \ =\ \sqrt{u_1u_2}\,, \qquad \tau_2 \ =\ \sqrt{u_5u_6}\,, \nonumber\\[5pt]
&&\rho_1 \ =\ -\frac{u_2}{1-u_3}\,, \qquad \rho_2 \ =\ -\frac{1-u_4}{u_5}\,, \nonumber\\[5pt]
&&\bar{\rho}_2 \ =\ -\frac{u_6 (1-u_3)}{u_1 (1-u_4)} \,, \qquad 1-\bar{\rho}_2 \ =\ \frac{1-u_7}{u_1 (1-u_4)}\,, \nonumber\\[5pt]
&&\bar{\rho}_1 - \bar{\rho}_2 \ =\ \frac{u_6 (1-u_7)}{u_1 (1-u_4)^2} \,,
\eea
which can be verified with the aid of eqs.~(\ref{eq:gletters}),
(\ref{eq:gsMRK}) and (\ref{eq:gsMRK2}).  Note that the matching is not for fully physical MRK kinematics, because $\bar\rho_i$ are not the complex conjugates of $\rho_i$ where we match.  However, it is on the correct sheet, and we can move freely to the region where $\bar\rho_i = \rho_i^*$ after integrating up the functions in MRK.

Thus, if we know the amplitudes on \CO{3}{4} on the physical sheet, we can then use that information as a boundary condition to obtain amplitudes in the MRK limit. The remaining question is how to go from the known boundary, which is the CO surface on the Euclidean sheet, to the CO surface on the physical sheet. The required analytic continuation~(\ref{eq:continuation}) is non-trivial due to the Gram determinant constraint~(\ref{eq:gramdet}); we cannot change $u_7$ by itself without also affecting other cross ratios. Also, $u_7$ is near 1 on \CO{3}{4} and it needs to be near 0 for the analytic continuation~(\ref{eq:continuation}). In the next section, we will describe a path that takes us to the correct Riemann sheet and connects the Euclidean CO surface all the way to MRK.


\section{The Long and Winding Road to MRK} \label{sec:longroad}

In this section we describe a sequence of connected paths that leads us to the MRK limit. Along the way we obtain analytic expressions for the amplitudes at each step in terms of $G$-functions. The expressions are rather bulky so we refrain from presenting them here. Finally in MRK, we obtain the results in terms of single-valued $\cG$-functions (see Appendix~\ref{app:polylog} and the ancillary files for this paper).  The overall sequence is depicted in Fig.~\ref{fig:longroad}.

\subsection{CO on the Euclidean sheet}

\begin{figure}[h]
\center
\includegraphics[width=0.6\linewidth]{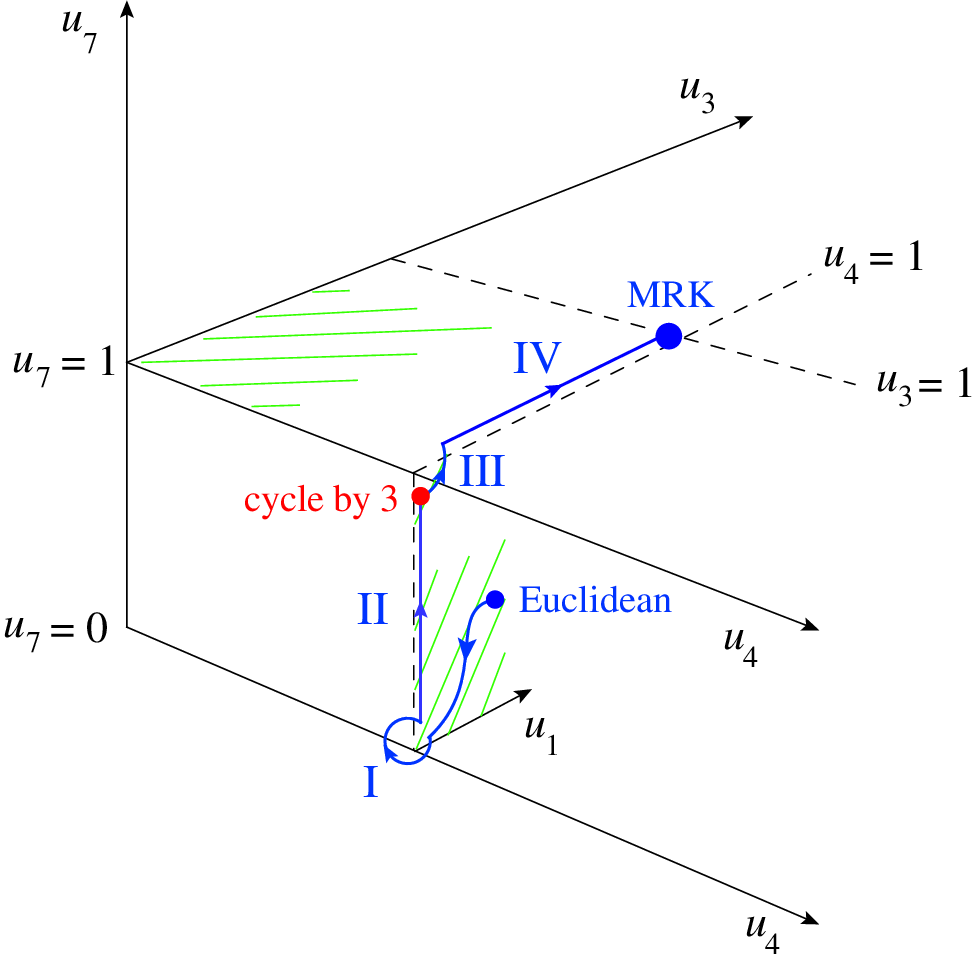}
\caption{A sequence of paths from the Euclidean CO surface to MRK.
It is a bit schematic in order to depict motion in four different
variables, $u_7,u_1,u_3,u_4$.
Starting from the Euclidean branch of the surface \CO{7}{1}, in step I
we analytically continue $u_7$ around its
complex origin, according to \eqn{eq:continuation}.
In step II we increase $u_7$ from 0 until it is close to 1,
but keeping $1-u_4 \ll 1-u_7$ on the physical branch of \CO{7}{1}.
In step III (the infinitesimal line), the hierarchy is reversed, so that
$1-u_7 \ll 1-u_4$, which places us on the physical branch of \CO{3}{4}.
In step IV, we increase $u_3$ from 0 until it is close to 1, at which point
we can match to MRK using \eqn{eq:MRKCOmatch}.}
\label{fig:longroad}
\end{figure}

Our starting point is \CO{7}{1} on the Euclidean sheet, which is labeled \textbf{\textcolor{blue}{Euclidean}} in Fig.~\ref{fig:longroad}. We choose the six free kinematic variables to be $(u_5,u_6,u_7,u_1,u_2,u_3)$, with $u_4$ fixed by~($\ref{eq:CO}$) to be close to unity. The amplitudes are known on this surface, and indeed the function space they belong to is very simple. Let us look at the symbol letters,
\be
\{u_5,\,u_6,\,u_7,\,1-u_7,\,u_1,\,1-u_1,\,u_2,\,u_3,\,1-u_4 \} \,.
\ee
This alphabet can be obtained by cyclically rotating $u_i\to u_{i+4}$ (mod 7) in \eqn{eq:COletters}, which provides the symbol alphabet on \CO{3}{4}. The only complicated letter is $1-u_4$. However, it drops out on the Euclidean sheet due to branch cut constraints~\cite{Dixon:2020cnr}. As a result, the function space factorizes as
\be\label{eq:COfunctions}
\left( \bigotimes_{i\in\{2,3,5,6\}}\left\{\ln^k(u_i),\ k\ge0\right\} \right) \otimes
\left( \bigotimes_{i\in\{7,1\}}\left\{ G(\vec{w};\, u_i),\ w_k\in\{0,1\} \right\} \right),
\ee
where the $G$ functions are defined in Appendix~\ref{app:polylog}. It is then very easy to perform the analytic continuation $u_7 \to e^{-2\pi i} u_7$ on the functions. For example, in a Lyndon basis for $G(\vec{w};\, u_7)$ with trailing 1's in $\vec{w}$ for all $G$-functions except $G(0;\, u_7) \equiv \ln u_7$, the only effect is to send $\ln u_7$ to $\ln u_7 -2\pi i $. In Fig.~\ref{fig:longroad}, the analytic continuation is represented as the tiny blue circle \textbf{\textcolor{blue}{I}} (which should really be pictured in the complex plane) around the origin $u_7=0$. This takes us to the correct physical Riemann sheet.

Now we want to pass on to \CO{3}{4}. First we go to the limit $(u_7,u_1) = (1,0)$ on \CO{7}{1}; this is along the blue line \textbf{\textcolor{blue}{II}} in Fig.~\ref{fig:longroad}. Schematically, the seven cross ratios at this point are $(u_1,\ldots,u_7) = (0,0,0,1,0,0,1)$, plus infinitesimal deviations. This looks exactly like the limit $(u_3,u_4) = (0,1)$ on \CO{3}{4}. However, there is an exchange-of-limits issue here. On \CO{7}{1} we have $1-u_4 \ll 1-u_7$; on the other hand, on \CO{3}{4} we have $1-u_4 \gg 1-u_7$. To see why the order of limits matters, let us look at a simple global function, $\Li_2(1-u_4 u_7)$. After analytically continuing $u_7$ according to \eqn{eq:continuation}, and then sending both $u_4,u_7\to1$, this function becomes $2\pi i \, \ln(1-u_4 u_7)$. Now, if we send $u_7\to1$ faster than $u_4$, the logarithm reduces to $\ln(1-u_4)$; on the other hand, if we send $u_4\to1$ faster, it reduces to $\ln(1-u_7)$. The result depends\footnote{Note that analytic continuation is essential in this simple example to see the order-of-limits issue. On the Euclidean sheet, the function $\Li_2(1-u_4 u_7)$ vanishes altogether in the $u_{4,7}\to1$ limit. As noted in Ref.~\cite{Dixon:2020cnr}, this is a general phenomenon for functions with the correct physical branch-cut behavior.} on whether $1-u_4 \ll 1-u_7$ or $1-u_4 \gg 1-u_7$. In the next subsection, we look at a path that connects these two different orders of limits.

\subsection{An infinitesimal line and CO on physical sheet}
We now need a path that connects the two limits $1-u_4 \ll 1-u_7$ and $1-u_4 \gg 1-u_7$, with all other cross ratios small. Such a line is ``infinitesimal'', in the sense that we stay near the same point in the space of the cross ratios; however, the two ends of this line approach the point $(u_4,u_7)=(1,1)$ from different angles in the $(u_4,u_7)$ plane. This is represented as the small arc or line \textbf{\textcolor{blue}{III}} in Fig.~\ref{fig:longroad}. As we learned from the above simple example $\Li_2(1-u_4 u_7)$, these angles matter since they change the limiting behavior of the function. To see what the kinematics looks like on this infinitesimal line, let us take the general scaling
\be
u_1,u_2,u_3,u_5,u_6, \, 1-u_4, \, 1-u_7 \sim \cO(\epsilon) , \quad \epsilon\to0 \,.
\ee
Solving the Gram determinant constraint~(\ref{eq:gramdet}) to leading order in $\epsilon$, we arrive at the following,
\begin{align}
\label{eq:IL}
\text{Infinitesimal line (IL):} \quad
\begin{cases}
& u_1,u_2,u_3,u_5,u_6, \, 1-u_4, \, 1-u_7 \ll 1, \quad \text{satisfying} \\
& (1-u_4)(1-u_7) = u_6 (1-u_4) + u_5 (1-u_7) \,.
\end{cases}
\end{align}

If we send the momentum $p_1\to0$ in formulas~(\ref{eq:uisi}) for the cross ratios $u_i$, we see that $(u_4,u_5,u_6,u_7) \to (1,0,0,1)$.  So the infinitesimal line is a special case of a soft limit of seven-point kinematics, where it approaches six-point kinematics.  In this case the six-point kinematics are near the ``origin'' since $(u_1,u_2,u_3) \to (0,0,0)$.  Using this observation, we have found a momentum twistor parametrization for this line, which is recorded in Appendix~\ref{app:IL} together with the limiting behavior of the $g$-letters. Indeed, we see that the letter $1-u_4 u_7$ appears in \eqns{eq:gsIL}{eq:gsIL2}. We can interpolate between the two $u_4,u_7\to1$ limits by sending $u_7\to 1$ or $u_4\to 1$ faster in this letter. One end of the line is \CO{7}{1}, the other end \CO{3}{4}, now on the physical Riemann sheet.

The function space the amplitudes belong to on this line is slightly more complicated, due to the letter $1-u_4 u_7$ in the alphabet~(\ref{eq:ILletters}) and the non-trivial constraint~(\ref{eq:IL}). If we choose $(u_1,u_2,u_3,u_4,u_5,u_7)$ as the independent variables (and $u_6$ becomes dependent on these), then a representation of the function space is\footnote{Such representations are in general not unique; they depend on the order in which one integrates over the independent variables. Given a choice of integration order, the set of $G$-functions one gets is called the \emph{fibration basis} \cite{Brown:2008um,Panzer:2014caa,Duhr:2019tlz}.}
\begin{multline}\label{eq:ILfunctions}
\left( \bigotimes_{i\in\{1,2,3\}}\left\{\ln^k(u_i),\ k\ge0\right\} \right) \otimes
\left\{\ln^k(1-u_4),\ k\ge0\right\} \\
\otimes \left\{G\left(\vec{w};\, \frac{u_5}{1-u_4}\right),\ w_k\in\{0,1\} \right\} \otimes
\left\{G\left(\vec{w};\, -\frac{1-u_7}{1-u_4}\right),\ w_k\in\{0,1\} \right\} \,.
\end{multline}
Alternatively, one can choose $u_7$ as the dependent variable; then the argument of the $G$-functions in the last factor becomes
$-(1-u_7)/(1-u_4) \to -u_6/(1-u_4-u_5)$.

Now that the final end of the infinitesimal line (\textbf{\textcolor{blue}{III}} in Fig.~\ref{fig:longroad}) touches \CO{3}{4} on the physical sheet, we can integrate from there to obtain amplitudes on \CO{3}{4}. The function space is more complicated than CO on the Euclidean sheet, since we now have to use the full alphabet~(\ref{eq:COletters}). Using $(u_1,u_2,u_3,u_4,u_5,u_6)$ as independent variables (with $u_7$ dependent), a representation of the function space is
\begin{multline}\label{eq:COu3u4functions}
\left( \bigotimes_{i\in\{1,2,5,6\}}\left\{\ln^k(u_i),\ k\ge0\right\} \right) \otimes
\left\{G(\vec{w};\, u_3),\ w_k\in\{0,1\} \right\} \\
\otimes
\left\{G(\vec{w};\, 1-u_4),\ w_k\in\left\{0,1,-\frac{(1-u_3)u_6}{u_1}\right\}
\right\} \,,
\end{multline}
where $1-u_7 = u_1(1-u_4)+u_6(1-u_3)$.

To reach the MRK limit, we bring $u_3$ from zero to one, along line \textbf{\textcolor{blue}{IV}} in Fig.~\ref{fig:longroad}.  In this limit, \eqn{eq:COu3u4functions} simplifies.  The second factor
becomes
\be
\left\{G(\vec{w};\, u_3),\ w_k\in\{0,1\} \right\}\ \to\
\left\{\ln^k(1-u_3),\ k\ge0\right\} \,.
\label{IVsimp2}
\ee
In the third factor, because $1-u_4 \ll 1$, the weight $w_k=1$ can be neglected, and then a rescaling can be performed, so that
\be
\left\{G(\vec{w};\, 1-u_4),\ w_k\in\left\{0,1,-\frac{(1-u_3)u_6}{u_1}\right\} \right\}
\ \to\
\left\{ G\left(\vec{w};\, -\frac{u_1(1-u_4)}{u_6(1-u_3)}\right),
\ w_k\in\{0,1\} \right\} \,.
\label{IVsimp3}
\ee
Comparing with \eqn{eq:MRKCOmatch}, we see that the only non-trivial $G$-function argument is $1/\bar{\rho}_2 = -u_1(1-u_4)/[u_6(1-u_3)]$, which is the only generic, order-one variable in \eqn{eq:MRKCOmatch}.  The generic letters
$\bar\rho_2$ and $1-\bar\rho_2$ in \eqn{eq:MRKCOmatch} correspond precisely to the function space~(\ref{IVsimp3}).  One can also invert the argument to $\bar\rho_2$, since this preserves the singular points $\{0,1,\infty\}$.  In the MRK labelling, the full function space at the matching point is
\begin{multline}\label{eq:MRKmatchfunctions}
\left( \bigotimes_{i\in\{1,2\}} \left\{\ln^k(\tau_i),\ k\ge0\right\} \right)
\otimes \left( \bigotimes_{i\in\{1,2\}}\left\{\ln^k(\rho_i),\ k\ge0\right\} \right)
\otimes \left\{ \ln^k (\bar\rho_1-\bar\rho_2), k\ge0 \right\} \\
\otimes \left\{G(\vec{w};\, \bar\rho_2),\ w_k\in\left\{0,1\right\} \right\} \,.
\end{multline}

\subsection{MRK}
We can now compute the amplitudes by integration at the MRK, with boundary conditions at $\rho_1\to0, \rho_2\to\infty, \bar{\rho}_1\to\bar{\rho}_2$ given by the point $u_{3,4}\to1$ on the physical sheet \CO{3}{4}, using the matching relations~(\ref{eq:MRKCOmatch}). The symbol letters are given in \eqn{eq:MRKletters}, and the function space is
\begin{multline}\label{eq:MRKfunctions}
\left( \bigotimes_{i\in\{1,2\}}\left\{\ln^k(\tau_i),\ k\ge0\right\} \right) \otimes
\bigg\{G(\vec{w};\, \rho_2),\ w_k\in\{0,1\} \bigg\} \otimes
\bigg\{G(\vec{w};\, \rho_1),\ w_k\in\{0,1,\rho_2\} \bigg\} \\
\otimes
\bigg\{G(\vec{w};\, \bar{\rho}_2),\ w_k\in\{0,1\} \bigg\} \otimes
\bigg\{G(\vec{w};\, \bar{\rho}_1),\ w_k\in\{0,1,\bar{\rho}_2\} \bigg\}  \,.
\end{multline}
Apart from the large logarithms $\ln\tau_{1,2}$, we see that the function space factorizes into a holomorphic part,
\be \label{eq:A2polylogs}
\bigg\{G(\vec{w};\, \rho_2),\ w_k\in\{0,1\}\bigg\} \otimes \bigg\{G(\vec{w};\, \rho_1),\ w_k\in\{0,1,\rho_2\}\bigg\} \,,
\ee
and its complex conjugate. The class of functions in \eqn{eq:A2polylogs} are sometimes referred to as \emph{$A_2$ polylogarithms} since their symbol letters $\{\rho_1,1-\rho_1,\rho_2,1-\rho_2,\rho_1-\rho_2\}$ correspond to the $\cA$-coordinates of an $A_2$ cluster algebra. Together with the anti-holomorphic part, we get two copies of the $A_2$ cluster algebra, and the class of functions is referred to as $A_2\times A_2$ polylogarithms.

One might have noticed an asymmetry in \eqn{eq:A2polylogs} between $\rho_1$ and $\rho_2$. When the last entry of the $G$-function is $\rho_1$, the front entries can include $\rho_2$, but not the other way around. This asymmetry stems from the order of integration, in that we choose to integrate over $\rho_1$ first and then $\rho_2$. Had we chosen to integrate over $\rho_2$ first, we would have obtained a different $G$-function representation of the same functions. These different representations are related by identities among polylogarithms, for example,
\be
G(\rho_2;\rho_1) = G(\rho_1;\rho_2) + \ln\rho_1 - \ln\rho_2 \pm i\pi \,.
\ee
Both sides represent the same simple function, $\ln(1-\rho_1/\rho_2)$, and the sign of $i\pi$ depends on the choice of branch for $\rho_2-\rho_1$.

The function space~(\ref{eq:MRKfunctions}) is further restricted by a \emph{first-entry} condition. On the Euclidean sheet, branch points can only occur when the cross ratios $u_i = g_{1i}$ become 0 or $\infty$. Bcause of this, the $u_i$ are the only first entries of the symbols of heptagon functions.  On the physical Riemann sheet, which is obtained by the analytic continuation~(\ref{eq:continuation}), the amplitude differs from the Euclidean version by discontinuities. The discontinuities can be obtained (at symbol level) by repeatedly clipping off $u_7 = g_{17}$ from the front of the symbol, thereby exposing later entries in the symbol.  Naively, after clipping off a string of $k$ successive $u_7$'s, one might expose arbitrary first entries for the symbols of the functions multiplying $(2\pi i)^k$ in the general MRK decomposition~(\ref{eq:ComponentExpansion}).  (For $k>2$, such terms are mixed with beyond-the symbol terms in $g$ and $h$ in \eqn{eq:ComponentExpansion}, but in principle they can be distinguished.) However, we have found, using a basis for the heptagon functions up through weight 6, that after clipping off a string of $k$ successive $u_7$'s, for $k<6$, the next entry is never $g_{4i}$, $g_{5i}$ or $g_{6i}$.

Inspecting \eqns{eq:gsMRK}{eq:gsMRK2} for the behavior of the $g$-letters in MRK, we see that the absence of the parity-odd letters $g_{5i}$ and $g_{6i}$ ensures that the exposed first symbol entries are only the following real combinations of $\rho_1,\rho_2$,
\be
|\rho_i|, \quad |1-\rho_i|, \quad |\rho_1-\rho_2| \,.
\label{eq:realrho}
\ee
This is sufficient to ensure that the resulting multiple polylogarithms are \emph{single-valued} (or \emph{real analytic}) functions of $\rho_1$ and $\rho_2$.

An analogous result for six-particle amplitudes and their MRK limit can be established by showing that clipping off a string of $k$ successive cross ratios $u=u_1$ never exposes any of the three parity-odd letters, $y_u$, $y_v$ or $y_w$. (These three letters are the only hexagon letters that don't reduce to the real combinations $|z|$ or $|1-z|$ in six-point MRK.) In the hexagon case, the result can be derived for arbitrary $k$, by induction on $k$, using the hexagon-function first-entry condition and the adjacent-pair relations~\cite{Caron-Huot:2019bsq}.  For example, one of the simplest adjacent-pair relations is $F^{y_u,u} = F^{u,y_u}$.  If this relation is used right after a string of $k$ $u$ entries in the symbol, and the left-hand side vanishes by the induction hypothesis, then the right-hand side must also vanish, corresponding to $k \to k+1$. Establishing that $y_v$ or $y_w$ cannot appear right after $k$ $u$'s is similar but a little more involved.\footnote{We thank A.~McLeod for discussions leading to this conclusion.}  Presumably a similar all-$k$ argument can be constructed in the seven-point case as well, but we won't do so here.

The first-entry constraint~(\ref{eq:realrho}) lifts to single-valued multiple polylogarithms.  They can be constructed using a single-valued map~\cite{DelDuca:2016lad} (see Appendix~\ref{app:polylog}).  We will denote them by single-valued $\cG$-functions. This further restricted, smaller function space that the MRK amplitudes belong to can be written as follows,
\be \label{eq:MRKfunctions2}
\left( \bigotimes_{i\in\{1,2\}}\left\{\ln^k(\tau_i),\ k\ge0\right\} \right) \otimes
\bigg\{\cG(\vec{w};\, \rho_2),\ w_k\in\{0,1\} \bigg\} \otimes
\bigg\{\cG(\vec{w};\, \rho_1),\ w_k\in\{0,1,\rho_2\} \bigg\} \,.
\ee
In principle, we know the derivatives with respect to $\rho_1,\rho_2$ as well as their complex conjugates $\bar{\rho}_1,\bar{\rho}_2$; we can integrate with respect to all of them in terms of $G$-functions, and then see that they come in combinations such that they can precisely be represented as the single-valued $\cG$-functions. We did so explicitly through weight four. Beyond weight four, we integrated only over $\rho_1,\rho_2$ to obtain the holomorphic part, and then used the single-valued map~(\ref{eq:svmap}) to obtain the full result in terms of $\cG$-functions.


\section{Comparison with Fourier-Mellin Construction} \label{sec:FM}
An all-loop formula in terms of a Fourier-Mellin integral for amplitudes in the MRK was already written down in \cite{Lipatov:2010qg} for the six-particle MHV amplitude at LLA, and was later generalized to sub-leading logarithmic accuracies, amplitudes with more legs, and beyond MHV \cite{Lipatov:2010ad,Bartels:2011ge,Lipatov:2012gk,Dixon:2012yy,Basso:2014pla,Dixon:2014iba,DelDuca:2016lad,DelDuca:2019tur}. For seven particles~\cite{Bartels:2011ge,Bartels:2013jna,Bartels:2014jya,DelDuca:2016lad}, the integral is \cite{DelDuca:2018hrv,DelDuca:2019tur}
\begin{multline} \label{eq:FMintegral}
\cR_{h_1,h_2,h_3}\, e^{i\delta_7} = 1 + 2\pi i \cdot \prod_{k=1,2}
\left[
\sum_{n_k=-\infty}^\infty \left(\frac{z_k}{\bar{z}_k}\right)^{\frac{n_k}{2}}
\int_{\cC_k} \frac{d\nu_k}{2\pi} \frac{|z_k|^{2i\nu_k} \tilde{\Phi}(\nu_k,n_k)}{(-\tau_k + i0)^{\omega(\nu_k,n_k)}}
\right] \times \\[7pt]
\bigg[
I^{h_1}(\nu_1,n_1) \cdot \tilde{C}^{h_2}(\nu_1,n_1,\nu_2,n_2) \cdot \bar{I}^{h_3}(\nu_2,n_2)
\bigg] \,,
\end{multline}
where $\omega, \tilde{\Phi}, I^h, \tilde{C}^{h}$ are known as the BFKL eigenvalue, impact factor, helicity flip kernel, and \emph{central emission vertex}, respectively. For example, to next-to-leading order, the positive helicity central emission vertex is \cite{DelDuca:2018hrv}
\begin{multline}
\tilde{C}^+ = \frac{\Gamma(1-i\nu_1-\frac{n_1}{2})
\Gamma(1+i\nu_2+\frac{n_2}{2})
\Gamma(i\nu_1-i\nu_2-\frac{n_1}{2}+\frac{n_2}{2})}
{g^2 \Gamma(i\nu_1-\frac{n_1}{2}) \Gamma(-i\nu_2+\frac{n_2}{2})
\Gamma(1-i\nu_1+i\nu_2-\frac{n_1}{2}+\frac{n_2}{2})} \ \times \\[7pt]
\bigg\{1 + g^2 \bigg[
D_1E_1 - D_2E_2 + E_1E_2 + \frac{1}{4}(N_1+N_2)^2 + V_1V_2 \\[5pt]
+ (V_1-V_2)(M-E_1-E_2) +
2\zeta_2 + i\pi(V_2-V_1-E_1-E_2)
\bigg] + \cO(g^4) \bigg\} \,,
\end{multline}
where
\bea
E_k &=& -\frac{1}{2}\frac{|n_k|}{\nu_k^2+\frac{n_k^2}{4}} +
\psi\left(1+i\nu_k+\frac{|n_k|}{2}\right) + 
\psi\left(1-i\nu_k+\frac{|n_k|}{2}\right) - 2\psi(1) \,, \nonumber\\
V_k &=& \frac{i\nu_k}{\nu_k^2+\frac{n_k^2}{4}} \,,\qquad
N_k = \frac{n_k}{\nu_k^2+\frac{n_k^2}{4}} \,,\qquad
D_k = -i\partial/\partial\nu_k \,, \nonumber\\
M &=& \psi\left(
 i\nu_1-i\nu_2-\frac{n_1}{2}+\frac{n_2}{2}\right) + \psi\left(
 1 - i\nu_1 + i\nu_2 - \frac{n_1}{2} + \frac{n_2}{2}
 \right) - 2\psi(1) \,.
\eea
We refer the readers to Ref.~\cite{DelDuca:2019tur} for the conjectured all-order expression for the central emission vertex. The other components in the integrand in (\ref{eq:FMintegral}) can be obtained by analytic continuation from the OPE limit, using the known finite-coupling OPE result \cite{Basso:2014pla}.

The integrand in \eqn{eq:FMintegral} has poles on the real axes at $\nu_1=\pi\Gcusp/4$, $\nu_2=-\pi\Gcusp/4$ and $\nu_1=\nu_2$,  corresponding to soft limits of the centrally produced gluons; and the prescription for the contours $\cC_1,\cC_2$ is for $\nu_1$ to go above $\pi\Gcusp/4$, $\nu_2$ to go below $-\pi\Gcusp/4$, and $\nu_1$ to go below $\nu_2$ (equivalently $\nu_2$ above $\nu_1$) \cite{DelDuca:2018hrv}. 

For $z_1 < 0$ and $z_2 > 0$, we can close the contour $\cC_1$ in the lower half complex plane, and $\cC_2$ in the upper half. After picking up poles on the real axes, we get
\begin{multline} \label{eq:FMdeformed}
\cR_{h_1,h_2,h_3}\, e^{i\delta_7}(z_1,z_2,\tau_1,\tau_2) =
\cR_{h_2,h_3}(z_2,\tau_2)\, e^{i\delta_6(z_2)}\, |z_1|^{i\pi\Gcusp/2} + \\
\cR_{h_1,h_2}(z_1,\tau_1)\, e^{i\delta_6(z_1)}\, |z_2|^{-i\pi\Gcusp/2} 
- \left|\frac{z_1}{z_2}\right|^{i\pi\Gcusp/2} +
\Sigma_{h_1,h_2,h_3}(z_1,z_2,\tau_1,\tau_2) \,,
\end{multline}
where $\cR_{h_i,h_j}$ is the six-gluon BDS-normalized amplitude, with $h_i,h_j$ the helicities of the centrally produced gluons, and
\be \label{eq:delta6}
\delta_6 (z) =
\frac{\pi \Gcusp}{4} \ln \frac{|z|^2}{|1-z|^4}
\ee
is the six-particle phase from the BDS ansatz (the analog of \eqn{eq:delta7} for the seven-particle phase). 
The first three terms in \eqn{eq:FMdeformed} are trivial; they are completely fixed by six-particle results. Genuinely new information to the seven-particle amplitudes is all contained in the last term, a sum over $\nu_1$ ($\nu_2$) residues in the lower (upper) half complex plane.
%
%
Denoting the seven-point Fourier-Mellin summand-integrand as
\begin{multline}
\mathcal{I}_{h_1,h_2,h_3}(\nu_1, n_1, \nu_2, n_2) = \prod_{k=1}^2 \left[\left(\frac{z_k}{\bar{z}_k}\right)^{\frac{n_k}{2}} \frac{|z_k|^{2 i \nu_k}\tilde{\Phi}(\nu_k, n_k)}{(-\tau_k + i 0)^{\omega(\nu_k, n_k)}}\right] \times\\
I^{h_1}(\nu_1, n_1) \tilde{C}^{h_2}(\nu_1, n_1, \nu_2, n_2) \bar{I}^{h_3}(\nu_2, n_2) \,,
\end{multline}
there are poles in $\mathcal{I}_{h_1,h_2,h_3}$ coming from explicit rational factors, gamma functions, and polygamma functions of the form $\psi^{(m)}(1\pm i\nu_k+\frac{|n_k|}{2})$. Combining the sum over residues with the Fourier-Mellin sum, we can arrange the non-trivial term in (\ref{eq:FMdeformed}) into the following form,
\begin{multline}
\label{eq:ressum}
\Sigma_{h_1,h_2,h_3}(z_1,z_2,\tau_1,\tau_2) = \sum_{j_1=1}^\infty \sum_{j_2 = -\infty}^{-1} \sum_{k_1=0}^{j_1} \sum_{k_2=-j_2}^0 \res_{\nu_i = -i \frac{j_i}{2}} \mathcal{I}_{h_1,h_2,h_3}(\nu_1, j_1 - 2 k_1, \nu_2, j_2 - 2 k_2) \\
= \sum_{j_1=1}^\infty \sum_{j_2 = -\infty}^{-1} \sum_{k_1=0}^{j_1} \sum_{k_2=-j_2}^0 c_{h_1,h_2,h_3}(j_1,k_1,j_2,k_2; \{\ln z_i, \ln \bar{z}_i, \ln \tau_i\}) \ z_1^{j_1-k_1} \bar{z}_1^{k_1} z_2^{j_2-k_2} \bar{z}_2^{k_2} \,,
\end{multline}
with some coefficient function $c_{h_1,h_2,h_3}$. We see that in the monomials above, $z_1,\bar{z}_1$ always come with non-negative powers, and $z_2,\bar{z}_2$ always come with non-positive powers, so the summation gives an expansion around $z_1=0$ and $z_2=\infty$. Note that the poles on the real $\nu_i$ axes (which in the small-coupling approximation all collapse to the origin of the Mellin space) would correspond to summands with $j_1 = 0$ or $j_2 = 0$. However, these terms are already taken care of by the contour deformation leading to~\eqref{eq:FMdeformed}; therefore they do not contribute to $\Sigma_{h_1,h_2,h_3}$ in \eqn{eq:ressum}.

Using the known and conjectured ingredients of the Fourier-Mellin integral, it is straightforward to calculate the necessary residues at any loop order. For example,
\begin{multline}
\Sigma_{+++}(z_1,z_2,\tau_1,\tau_2) = g^2\left(\frac{z_1}{z_2}\right) +
g^4 \left(\frac{z_1}{z_2}\right)\biggl[
  \ln(-\tau_1)(1+\ln|z_1|^2) + \ln(-\tau_2)(1+\ln|z_2|^2) \\[10pt]
- \frac{\ln|z_1|^2(\ln|z_1|^2-3) + \ln|z_2|^2(\ln|z_2|^2-3)}{2}
- \frac{\pi^2}{3} - 4
\biggl]
\ +\ \cO(g^6, z_1^2, 1/z_2^2, \bar{z}_1, 1/\bar{z}_2) \,.
\end{multline}
We can then compare the resulting series expansions with the bootstrapped amplitudes, which we have also expanded around the same small $z_1$, large $z_2$ limit\footnote{From \eqn{eq:simplicial}, this limit is equivalent to expanding around small $\rho_1$ and large $\rho_2$.} with $\bar{z}_i$ the complex conjugates of $z_i$. We have performed the checks with the full expansion (\ref{eq:ressum}) through all terms with $|j_1| + |j_2| \leq 5$.  We have also checked the holomorphic part, by ignoring any $\bar{z}_i$ dependence, which corresponds to the special case in \eqn{eq:ressum} with $k_1,k_2=0$. This case is less computationally intensive, and so we could check all terms with $|j_i| \leq 40$ up to three loops and with $|j_i| \leq 10$ at four loops. We find perfect agreement between the Fourier-Mellin sum-integral and the bootstrapped amplitudes.


\section{A One-dimensional Line}  \label{sec:line}

An important advantage of knowing the amplitude at the function level, in contrast to knowing only its symbol, is that it can readily be evaluated numerically. To illustrate our results, in this section we provide details on the numerical evaluation of the components of the amplitude according to the decomposition~\eqref{eq:ComponentExpansion}. Of course, in generic multi-Regge kinematics the amplitude depends on the two complex variables $z_1, z_2$ and the small parameters $\tau_1, \tau_2$ defined in terms of cross ratios in \eqn{eq:MRKvariables}. Hence, if we want to plot the kinematic dependence of the amplitude we need to specialize the kinematics even further, say to some one-dimensional line. In principle, every line through the $(z_1,z_2)$ space is equally adequate. However, we will investigate a particularly symmetric line, defined by
\be
    \overline{\rho}_1 = \rho_1 = \frac{1}{\rho_2} = \frac{1}{\overline{\rho}_2}
\ee
in terms of the simplicial coordinates~\eqref{eq:simplicial} and their complex conjugates. This line is invariant under the parity and target-projectile symmetry transformations described in Appendix~\ref{app:MRKsym}. We will therefore refer to it as the target-projectile-parity (TPP) symmetric line. 

On the TPP symmetric line, the MRK symbol alphabet~\eqref{eq:MRKletters} simplifies to
\be
    \{\tau_1, \tau_2, \rho_1, 1-\rho_1, 1+\rho_1\}. 
\ee
Hence, on this line the component functions $g_{h_1,h_2,h_3}^{l;n_1,n_2}$, $h_{h_1,h_2,h_3}^{l;n_1,n_2}$ may be described solely in terms of harmonic polylogarithms (HPLs) \cite{Remiddi:1999ew}, $H_{\vec{w}}(\rho_1)$ with $w_k \in \{0,1,-1\}$. We provide the explicit expressions of all component functions in terms of HPLs in the ancillary file {\tt g\_h\_TPS.m}.

For example, the leading-log components of the four-loop MHV amplitude on the TPP symmetric line take the form
\begin{align}
    \left.g_{+++}^{4;0,3}\right|_\text{TPP} &= 8\bigg[ \frac{1}{3}H_0^3 H_1 + \frac{1}{2}H_0^2 \left(7H_1^2 + 2H_2\right) + H_0\left(8H_1^3 + 5 H_1 H_2 - H_3 - 3 H_{2,1}\right)\notag\\
    &\qquad +4 H_1^4 -H_1 (5 H_3 + 3 \zeta_3)\bigg],\notag\\
  \left.  g_{+++}^{4;1,2}\right|_\text{TPP} &= 8\bigg\{ H_0^3 (H_1 - H_{-1}) -\frac{1}{2} H_0^2(2H_{-2}+ H_{-1}^2 +9 H_{-1}H_1 -17 H_1^2 -2 H_2) \notag\\
  &\qquad + H_0 \Big[ H_{-3} - 2 H_{-1}H_1^2 + 12 H_1^3 + \frac{1}{2} H_{-1} (H_2 - 2 H_{-2})\notag\\
  &\qquad \qquad -\frac{1}{2}H_1 (3 H_{-2}+H_{-1}^2 - 6 H_2) - H_3 + 5 (H_{-2,1}+H_{2,-1}) - 9 H_{2,1}\Big] \notag\\
  & \qquad+ \frac{1}{2}H_1 \Big[ 3 H_{-3}- 6 H_3 - 2 H_{-2,-1} + 8 (H_{-2,1}+H_{2,-1}-2H_{2,1}) - \zeta_3 \Big] \notag\\
  & \qquad-\frac{1}{2} H_{-1}(H_3-2H_{-3} -2H_{-2,1} -2 H_{2,-1}+8 H_{2,1} -\zeta_3) + 4 H_1^4\bigg\},
\end{align}
for the $g$-components, as well as
\begin{align}
  \left.h_{+++}^{4;0,2}\right|_\text{TPP} &= \frac{16}{3}H_0^3 H_1 + 2H_0^2 (15 H_1^2 + 4 H_2) + \frac{4}{3} H_0 (32 H_1^3 + 15 H_1 H_2 - 6 H_3 - 9 H_{2,1})\notag\\
  & \qquad  + 4 H_1 (4 H_1^3 - 5 H_3 - 3 \zeta_3),\notag\\
  \left.h_{+++}^{4;1,1}\right|_\text{TPP} &= 8 \bigg\{ H_0^3 (H_1 - H_{-1}) + \frac{1}{2}H_0^2 \Big[2H_2 + 10 H_1^2 - 3H_{-1}H_1 -2 H_{-2}\Big] \notag\\
  &\qquad + \frac{1}{6} H_0 \Big[6 H_{-3} - 15 H_{-2}H_1 + 38 H_1^3 + 24 H_1 H_2 - 6 H_3\notag\\
  &\qquad \qquad -3 H_{-1}(H_1^2+H_2)  + 18(H_{-2,1}+H_{2,-1})-30 H_{2,1}\Big] \notag\\
  &\qquad + \frac{1}{2} H_1\Big[5 H_{-3} + 4 H_1^3 - 8 H_3 + 2(H_{-2,1}+H_{2,-1} - 2 H_{2,1})-\zeta_3\Big] \notag\\
  &\qquad + \frac{1}{2} H_{-1} (H_3-2 H_{2,1}+\zeta_3)\bigg\},
\end{align}
for the $h$-components, where we use the compact notation~\eqref{eq:CompactHPL}~\cite{Remiddi:1999ew}, and all HPLs are evaluated at $\rho_1$. We observe that at all loop orders $L=2,3,4$ the LLA coefficient functions of the MHV amplitude with all powers of the same logarithm, $g_{+++}^{L;0,L-1}$ and $h_{+++}^{L;0,L-2}$, are given by multiple polylogarithms with letters $\{0,1\}$ only. The NMHV amplitudes do not share this property.

\begin{figure}[h]
\center
\begin{tabular}{cc}
\includegraphics[width=0.48\linewidth]{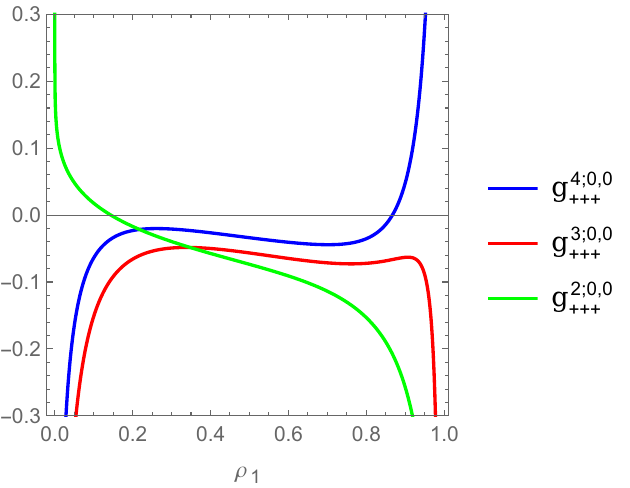}&
\includegraphics[width=0.48\linewidth]{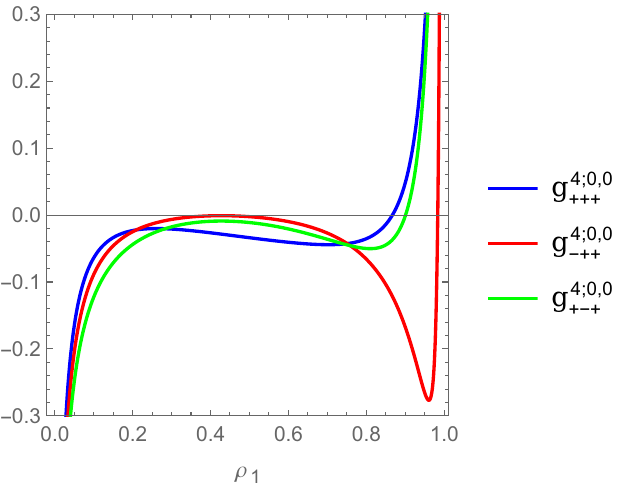}\\ \hspace{-1.6cm}(a) & \hspace{-1.6cm}(b)
\end{tabular}
\caption{Plots on the TPP symmetric line of the maximally subleading component in the LLA expansion of the amplitude, i.e.~the $n_1=n_2=0$ component without any logs of the small variables $\tau_1,\tau_2$. (a) shows the 2-, 3- and 4-loop component of the MHV amplitude. (b) shows a comparison between the 4-loop maximally subleading components of the MHV, $({-}{+}{+})$ NMHV and $({+}{-}{+})$ NMHV component respectively. To facilitate numerical comparisons, the $L$-loop components are divided by $(-16)^L$. }
\label{fig:MaximallySubleading}
\end{figure}

In Fig.~\ref{fig:MaximallySubleading}, we plot different loop orders of the maximally subleading logarithmic terms in the MHV amplitude, as well as the four-loop contributions to the MHV and NMHV $({-}{+}{+})$ and $({+}{-}{+})$ amplitudes respectively. For conciseness, we only show the maximally subleading real component in the LLA expansion. 
On the TPP line we have
\begin{align}
z_1 = \frac{\rho_1}{1+\rho_1}, \quad z_2 = \frac{1+\rho_1}{\rho_1}.
\end{align} 
In the limit $\rho_1\rightarrow 0$, we therefore have
\begin{align}
z_1 \rightarrow 0, \qquad z_2 \rightarrow \infty
\end{align}
with $z_1 z_2 = 1$ fixed. This corresponds to a double soft limit in which the seven-point BDS-normalized amplitude goes smoothly into the five-point one, which is identically $1$.  Inspecting \eqn{eq:ComponentExpansion}, $\cR_{h_1,h_2,h_3}\to1$, and so the (divergent) behavior of the $g$-components is captured entirely by the behavior of the $e^{i \delta_7}$ phase, where $\delta_7$ is given in \eqn{eq:delta7}.  Since $\delta_7$ does not contain $\ln\tau_{i}$, all components multiplying these logarithms vanish in the $\rho_1\rightarrow 0$ limit.  On the other hand, the divergent behavior of $g_{h_1,h_2,h_3}^{L;0,0}$ in Fig.~\ref{fig:MaximallySubleading} as $\rho_1\to0$ agrees with the double-soft prediction obtained by expanding $e^{i\delta_7}$ perturbatively.

In the limit $\rho_1 \rightarrow 1$, the $z_i$ variables go to
\begin{align}
z_1 \rightarrow \frac{1}{2} \qquad z_2 \rightarrow 2.
\end{align}
This does not look like a particularly singular point. However, this limit approaches the point $(\rho_1,\rho_2)=(1,1)$ in the full space of simplicial coordinates. Away from the TPP line, the limit $\rho_1\rightarrow \rho_2$ corresponds to a soft limit that should go smoothly into the six-point MRK amplitude evaluated at $z = -z_1 z_2 = +\rho_2$~\cite{DelDuca:2018hrv}. Then, approaching the point $(\rho_1, \rho_2)=(1,1)$ corresponds to the singular $z\rightarrow 1$ limit of the six-point amplitude. 
Indeed, we find that the log-divergent parts in the $\rho_1\rightarrow 1$ limit on the TPP line agree with the corresponding terms in the $z\rightarrow 1$ limit of the respective six-point amplitude, taking into account also the behavior of $\delta_7$. However, there is a mismatch that does not depend on $\rho_1$ but only on the small logarithms $\ln \tau_1$, $\ln \tau_2$.  This mismatch arises due to an exchange-of-limits issue:  The TPP line approaching the point $(1,1)$ from a different direction in the $(\rho_1,\rho_2)$ plane ($\rho_1=1/\rho_2$) than the direction corresponding to the soft limit ($\rho_1=\rho_2$).
 
\begin{figure}[h]
\center
\begin{tabular}{cc}
\includegraphics[width=0.48\linewidth]{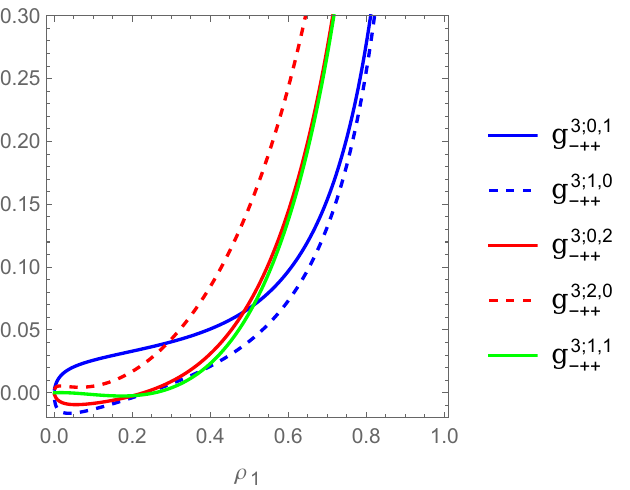}
&\includegraphics[width=0.48\linewidth]{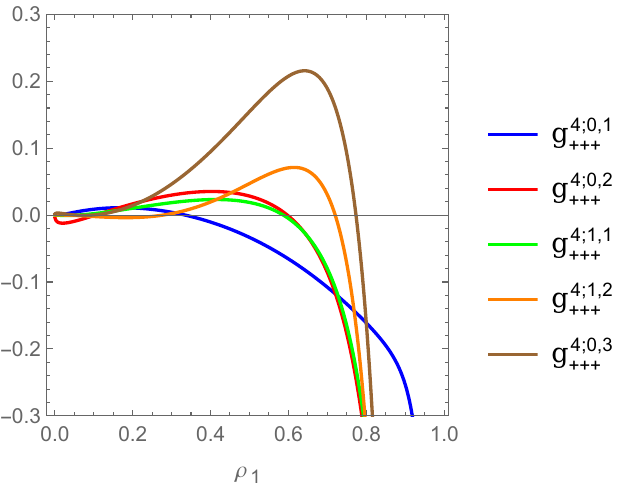}\\ \hspace{-1.6cm}(a) & \hspace{-1.6cm}(b)
\end{tabular}

\caption{Plots on the TPP symmetric line of different terms in the leading-log approximation (LLA). (a) shows the different components of the 3-loop NMHV $({-}{+}{+})$ amplitude related by the target-projectile exchange. (b) shows different components of the 4-loop MHV amplitude. The omitted components are related by target-projectile symmetry. To make the components of similar numerical size, we divided them by $(-16)^L/(-2\pi)^l$, where $l$ is the total order in the LLA expansion, $l=n_1+n_2$.}
\label{fig:LLAComponents}
\end{figure}

On the TPP symmetric line, the flip along the beam axis~\eqref{eq:cFmap} reduces to an exchange of $\tau_1$ and $\tau_2$. Hence, we have
\begin{align}
g_{h_1,h_2,h_3}^{L; n_1, n_2}\big|_\text{TPP} &= g_{h_3,h_2,h_1}^{L; n_2, n_1}\big|_\text{TPP}, \qquad 
h_{h_1,h_2,h_3}^{L; n_1, n_2}\big|_\text{TPP} = h_{h_3,h_2,h_1}^{L; n_2, n_1}\big|_\text{TPP}.
\end{align}
Accordingly, the components of the MHV and $({+}{-}{+})$ NMHV amplitude are invariant under the $n_1 \leftrightarrow n_2$ exchange. However, this is not the case for the components of the $({-}{+}{+})$ amplitude, which are mapped to the $({+}{+}{-})$ amplitude instead. To illustrate the difference between components of the $({-}{+}{+})$ amplitude that are related by the $n_1 \leftrightarrow n_2$ exchange, we plot them side by side at three loops in Fig.~\ref{fig:LLAComponents}(a) with continuous and dashed lines respectively.

Finally, in Fig.~\ref{fig:LLAComponents}(b) we give all independent components in the LLA expansion of the four-loop MHV amplitude that multiply large logarithms $\ln \tau_i$. As required by the double soft limit onto the five-point amplitude, they all vanish in the limit $\rho_1\rightarrow 0$. 

Notably, the components with the same overall order in the LLA expansion (i.e.~with the same $n_1+n_2$) have the same asymptotic behavior in the $\rho_1 \rightarrow 1$ limit and differ only by certain constants. Again, if we had approached the point $(1,1)$ in the space of the simplicial coordinates from the direction corresponding to the soft limit onto the singular $z\rightarrow 1$ limit of the six-point amplitude, the different components would have agreed exactly without any additional constants.


\section{Conclusions}
In this paper, we obtained two-to-five gluon amplitudes in planar $\cN=4$ SYM, in the ``long cut'' MRK where all centrally produced gluons have negative energies. We made use of iterated coproducts (derivatives) of the amplitudes obtained in earlier work, which are valid for general kinematics. We then integrated up these derivatives along a series of paths that connects a boundary on the Euclidean sheet, where the amplitudes are known in closed form, to MRK on the physical Riemann sheet.  In this way, we obtained the amplitudes in MRK through four loops and to all logarithmic accuracy. 

We compared our results to the BFKL approach in which amplitudes in the MRK are given by certain Fourier-Mellin sum-integrals. Our results are consistent with a recent all-order conjecture for the central emission vertex, and provide a stringent test of that conjecture.

The techniques we employed to obtain the amplitudes are quite general, and are commonly utilized in the amplitude bootstrap program. In the amplitude bootstrap, one usually starts with some number of unknown coefficients in an ansatz for the amplitude, which is often expressed in terms of its coproducts. One then finds paths that connect among different limiting kinematics; often times the limiting behaviors in these kinematics are known, based on physical grounds.  For example, the (BDS-normalized) amplitude should reduce to the case with fewer legs when some particles are taken to be soft or collinear. These known limiting behaviors then help to fix the initial unknown coefficients.

In this paper we turned this technique on its head. Since all coproducts of the amplitudes have been fixed completely in previous work, all we have to do is find a path that connects some known limiting kinematics to the kinematics of interest, which is MRK in this case. We can then just integrate along this path to obtain amplitudes in MRK.

The MRK limit we looked at is in the Mandelstam region where the only non-trivial Regge cut at seven points contributes. There are other Regge cuts, which are considered trivial in the sense that they are simply six-particle cuts, and are therefore not new for seven-particle amplitudes.  Consistency checks between the bootstrapped amplitudes and the BFKL approach would still be provided by computing the seven-point amplitudes in these other Mandelstam regions, using the techniques described here.  Various physical ``self-crossing'' kinematics could be explored similarly~\cite{Dixon:2016epj}.  However, we leave such computations for future work.


\acknowledgments
We are grateful to Andrew McLeod for useful discussions.
This research was supported by the US Department of Energy under
contract DE--AC02--76SF00515.
YL acknowledges support from the Benchmark Stanford Graduate Fellowship. 
JM is supported by the International Max Planck Research School for Mathematical and Physical Aspects of Gravitation, Cosmology and Quantum Field Theory. 


\appendix
\section{BDS ansatz and BDS-normalized amplitudes}
\label{app:bds}
The BDS ansatz \cite{Bern:2005iz} captures the infrared divergences of amplitudes in planar $\cN=4$ SYM. In dimensional regularization with the planar coupling $g^2 = {g_{\text{YM}}^{2}N_c}/{(16 \pi^{2})}$, the $n$-particle BDS ansatz is
\be
\cA_n^{\text{BDS}} = \cA_n^{\text{MHV,tree}}\cdot \exp\left[\sum_{L=1}^\infty g^{2L}\left(f^{(L)}(\epsilon)\cdot M_n(L\epsilon)\, +\, \text{const.} \right) \right],
\label{BDSansatz}
\ee
where the MHV tree super-amplitude is
\be
\cA_n^{\text{MHV,tree}}\ =\ \frac{\delta^8(\sum_{i=1}^n \lambda_i \eta_i)}
   {\langle12\rangle \langle23\rangle \cdots \langle n1\rangle} \,,
\label{treeMHVsuper}
\ee
and the function $f(\epsilon)$ starts off as\footnote{Subleading terms in $\epsilon$ are also important in $f(\epsilon)$, due to the second-order poles in $M_n$; however, we are ignoring them here since they are irrelevant for our purposes.}
\be
f(\epsilon) = \sum_{L=1}^\infty g^{2L} f^{(L)}(\epsilon) = \frac{\Gcusp}{4} \, + \, \cO(\epsilon).
\ee
The cusp anomalous dimension $\Gcusp$ is known to all orders in planar $\cN=4$ SYM~\cite{Beisert:2006ez},
\be\label{eq:Gcusp}
\frac{\Gcusp}{4} = g^2 - 2\,\zeta_2\,g^4 + 22\,\zeta_4\,g^6 - \left(219\,\zeta_6 + 8\,\zeta_3^2\right) g^8 + \cO(g^{10})\,.
\ee

The quantity $M_n$ is the one-loop amplitude normalized by the tree amplitude.
In the case of seven particles~\cite{Bern:1994zx,Dixon:2016nkn},
\bea \label{eq:M7}
M_7(\epsilon)\ =\ \sum_{i=1}^{7} &&\biggl[ -\frac{1}{\epsilon^{2}}\left( \frac{\mu^{2}}{-s_{i,i+1}} \right)^{\epsilon} +
 \Li_{2} \left(1{-}\frac{1}{u_{i}} \right)
 + \frac{1}{2} \ln\left(\frac{u_{i+2}u_{i{-}2}}{u_{i+3}u_{i}u_{i{-}3}}\right)
               \ln u_{i} 
 \nonumber\\
  &&\null  
 + \ln s_{i,i+1} \,
   \ln\left( \frac{s_{i,i+1}s_{i+3,i+4}}{s_{i+1,i+2}s_{i+2,i+3}} \right)
  + \frac{3}{2} \zeta_{2} \biggr] + 
\cO(\epsilon) .
\eea
\vspace{5pt}
The \emph{BDS-normalized amplitude} is then defined by
\be
\cR_n \equiv \frac{\cA_n}{\cA_n^{\text{BDS}}} \,,
\ee
which admits the N$^k$MHV decomposition $\cR_n = \cR_n^{\text{MHV}} + \cR_n^{\text{NMHV}} + \dots$.

\section{BDS-like ansatz and BDS-like normalized amplitudes}
\label{app:bdslike}

In the bootstrap program, it is useful to factor out another finite quantity from the full amplitude. We define the BDS-like ansatz \cite{Alday:2009dv,Caron-Huot:2016owq,Dixon:2016nkn},
\be\label{eq:bdslike}
\cA_n^{\text{BDS-like}} \equiv \cA_n^{\text{BDS}}
\exp\left[ -\frac{\Gcusp}{4} \, \cE_n^{(1)} \right] \,,
\ee
where for $n=7$,
\be
\label{eq:EMHV1}
\cE_7^{(1)} = \sum_{i=1}^7 \biggl[ \Li_2\left(1-\frac{1}{u_i}\right)
  + \frac{1}{2} \ln \left(\frac{u_{i+2}u_{i{-}2}}{u_{i+3}u_{i}u_{i{-}3}}\right)
               \ln u_i \biggr].
\ee
We see from formula~(\ref{eq:M7}) that the additional factor $\exp\left[ -\frac{\Gcusp}{4} \, \cE_7^{(1)} \right]$ removes all dependence on three-particle Mandelstam variables from the seven-point BDS ansatz~(\ref{BDSansatz}), so that BDS-like normalized amplitudes obey the three-particle Steinmann relations. This property allows them to be described with a much smaller function space of polylogarithms, which also obey \emph{cluster adjacency}~\cite{Drummond:2017ssj,Drummond:2018dfd,Drummond:2018caf} or \emph{extended Steinmann relations}~\cite{Caron-Huot:2019bsq}.

We then define \emph{BDS-like normalized} amplitudes ($\cE_7$ for MHV and $E_7$ for NMHV), now specialized to $n=7$,
\begin{align}
\label{eq:bdslikeampmhv}
\cE_7 &\equiv \frac{\cA_7^{\text{MHV}}}{\cA_7^{\text{BDS-like}}}
= \cR_7^{\text{MHV}} \, \exp\left[\frac{\Gcusp}{4} \, \cE_7^{(1)}\right] \,,
\\[7pt]
E_7 &\equiv \frac{\cA_7^{\text{NMHV}}}{\cA_7^{\text{BDS-like}}} 
= \cR_7^{\text{NMHV}} \, \exp\left[\frac{\Gcusp}{4} \, \cE_7^{(1)}\right] \,,
\label{eq:bdslikeampnmhv}
\end{align}
where we recall that $\cR_7$ is the BDS-normalized amplitude $\cA_7/\cA_7^{\text{BDS}}$. Since $\cR_7^{\text{MHV}}$ starts at two loops, we see that $\cE_7^{(1)}$ in \eqn{eq:EMHV1} is indeed the one-loop MHV BDS-like normalized amplitude, as befits its notation.

In Ref.~\cite{Dixon:2020cnr} the iterated coproducts for $\cE_7$ and all the components of $E_7$ were provided in terms of lower-weight functions, as well as their values on the Euclidean CO surface.  So it is $\cE_7$ and $E_7$ that we actually transport to the MRK limit.  At that point, one can use \eqns{eq:bdslikeampmhv}{eq:bdslikeampnmhv} to convert the results to those for BDS normalization.


\section{Symmetries of MRK}
\label{app:MRKsym}

Here we record a few discrete symmetry transformations for gluon scattering in the MRK limit \cite{DelDuca:2016lad}. First there is the parity transformation $\cP$; this reverses the gluon helicities,
\be
\cP \cdot \cR_{h_1, h_2, h_3} = \cR_{-h_1, -h_2, -h_3} \,,
\ee
recalling that $h_1,h_2,h_3$ are helicities of the centrally produced gluons. In terms of kinematic variables, it fixes $\tau_i$ and sends $\rho_i$ to their complex conjugates,
\be
\cP : \quad \rho_i \leftrightarrow \bar{\rho}_i \,.
\ee

The second symmetry $\cF$ is a flip along the collision beam axis,
\be
\label{eq:cFmap}
\cF \cdot \cR_{h_1, h_2, h_3} = \cR_{h_3, h_2, h_1} \,.
\ee
In terms of kinematic variables, it acts as,
\be
\cF : \quad
\tau_1 \leftrightarrow \tau_2 \,, \quad \rho_1 \leftrightarrow \frac{1}{\bar{\rho}_2} \,.
\ee
The combination of the above two symmetries is often referred to as the \emph{target-projectile} symmetry,
\be
\textrm{\emph{target-projectile}} \equiv \cF \circ \cP : \quad
\tau_1 \leftrightarrow \tau_2 \,, \quad \rho_1 \leftrightarrow \frac{1}{\rho_2} \,.
\ee


\section{Multiple polylogarithms}
\label{app:polylog}

Here we describe explicit representations of multiple polylogarithms \cite{Goncharov:1998kja,Goncharov:2001iea}. We will often refer to these as $G$-functions; they are defined recursively via the integral
\be\label{eq:GFunc}
G(\vec{w};z) \equiv G(w_1,w_2,\dots,w_n; z) \equiv \int_0^z \frac{dt}{t-w_1} G(w_2,\dots,w_n; t),
\ee
with the base case $G(; z) \equiv 1$. The special case where $w_1,\dots,w_n$ are all 0 is defined to be
\be
G(0,\dots,0; z) \equiv \frac{1}{n!}\ln^n z.
\ee
We refer the readers to the review \cite{Duhr:2014woa} for properties of the $G$-functions and to Ref.~\cite{Duhr:2019tlz} for a computer package, {\sc PolyLogTools}, that can deal with explicit computations.

Simply by looking at \eqn{eq:GFunc}, we see that $G$-functions carry many (iterated) branch points. However, certain combinations of $G$-functions, whose arguments are complex variables and their conjugates, are free of branch points, and are therefore real-analytic or single-valued functions; we will call these \emph{single-valued} multiple polylogarithms. Some explicit examples are images of the single-valued map defined in Ref.~\cite{DelDuca:2016lad},
\be \label{eq:svmap}
\mathbf{s}: G(\vec{w}; z) \mapsto \cG(\vec{w}; z) \,.
\ee
where the $\cG$-function is a combination of $G$-functions with arguments the holomorphic variables $w_i,z$ or their complex conjugates $\bar{w}_i,\bar{z}$. An example is the single-valued logarithm,
\be
\mathbf{s}(\ln z) = \ln |z|^2 = \ln z + \ln \bar{z} \,.
\ee
The single-valued map is implemented in {\sc PolyLogTools}; however, the output functions are not always in a convenient basis. For the readers' convenience, we provide an ancillary file {\tt cGToG.m} that converts the $\cG$-functions we need in \eqn{eq:MRKfunctions2} into combinations of $G$-functions, where the only functions that show up are the following,
\begin{align}
\begin{cases}
G(\vec{w}; \rho_1),       & w_i \in \{0,1,\rho_2\} \,, \\
G(\vec{w}; \bar{\rho}_1), & w_i \in \{0,1,\bar{\rho}_2\} \,, \\
G(\vec{w}; 1/\rho_2),       & w_i \in \{0,1\} \,, \\
G(\vec{w}; 1/\bar{\rho}_2), & w_i \in \{0,1\} \,,
\end{cases}
\end{align}
and in addition, the $\vec{w}$ are always Lyndon words. The above set of $G$-functions are convenient in the sense that they can be easily expanded around $\rho_1,\bar{\rho}_1=0$ and $\rho_2,\bar{\rho}_2=\infty$ using {\sc PolyLogTools}, thus making it easy to compare with the Fourier-Mellin residue sum described in Sec. \ref{sec:FM}.

A particularly simple subset of $G$-functions are harmonic polylogarithms (HPLs)~\cite{Remiddi:1999ew}. They encompass all $G$-functions with indices (or weight vector components) $w_i$ taken only from the alphabet $\{0,1,-1\}$. Excluding the cases where the rightmost entry is zero, they can be compactly expressed via the notation
\begin{align}
\label{eq:CompactHPL}
H_{n_1,n_2,...}(z) = \sigma \times G(\underbrace{0,...,0,\text{sign}(n_1)}_{n_1 \text{ entries}}, \underbrace{0,...,0,\text{sign}(n_2)}_{n_2 \text{ entries}}, ...; z). 
\end{align}
where the sign $\sigma$ is $1$ ($-1$) if the number of ``1'' entries
in the $G$ index list is even (odd). For example,
\begin{align}
H_{-2,3}(z) = - G(0,-1,0,0,1;z).
\end{align}


\section{Symbol alphabet in different kinematic limits}
\label{sec:letterlimits}

In this appendix we describe what the 42 symbol letters for heptagon functions become in the various limits we need.  The limiting values of the letters provide a complete description of the derivatives of all heptagon functions in the necessary kinematic regions.

\subsection{CO surface}

To obtain a twistor parametrization for the CO surface~(\ref{eq:CO}), we start with the OPE parametrization~\cite{Basso:2013aha},
\be
\label{eq:ZsOPE}
(Z_1,Z_2,\ldots,Z_7) =
\begin{pmatrix}
\frac{S_1}{\sqrt{F_1}}  & 1 & -1 & -S_2\sqrt{F_2} & 0 & 0 & 0 \\
0 & 0 & 0 & \frac{1}{T_2\sqrt{F_2}} & \frac{S_2+T_2F_2}{T_2S_2\sqrt{F_2}}
        & 1 & \frac{1}{S_1\sqrt{F_1}} \\
\frac{\sqrt{F_1}}{T_1} & 0 & 0 & -\frac{1}{T_2\sqrt{F_2}}
        & -\frac{1}{T_2\sqrt{F_2}} & 0 & \frac{\sqrt{F_1}}{T_1} \\
T_1\sqrt{F_1} & 0 & 1 & \frac{1+T_2S_2F_2+T_2^2}{T_2\sqrt{F_2}}
        & \frac{1}{T_2\sqrt{F_2}} & 0 & 0
\end{pmatrix} \,,
\ee
where
\be
T_j = e^{-\tau_j}, \qquad
S_j = e^{\sigma_j}, \qquad
F_j = e^{i\phi_j} \,,
\ee
$j=1,2$, and then scale
\be
T_j \mapsto T_j \cdot \epsilon,  \qquad
S_j \mapsto \frac{S_j}{\epsilon}, \qquad
F_j \mapsto \frac{F_j}{\epsilon^2} ,
\ee
with $\epsilon\to0$. Plugging these into the $g$-letters~(\ref{eq:gletters}), they collapse to the following,
\vspace{5pt}
\begin{align}
g_{11} &= u_1,& g_{21} &= 1,&
g_{31} &= 1,& g_{41} &= 1-u_4,&
\nonumber\\[1ex]
g_{12} &= u_2,& g_{22} &= 1,&
g_{32} &= 1-u_4,& g_{42} &= 1,&
\nonumber\\[1ex]
g_{13} &= u_3,& g_{23} &= 1-u_3,&
g_{33} &= 1,& g_{43} &=1-u_4,&
\nonumber\\[1ex]
g_{14} &= u_4,& g_{24} &= 1-u_4,&
g_{34} &= 1,& g_{44} &= 1-u_3,&
\nonumber\\[1ex]
g_{15} &= u_5,& g_{25} &= 1,&
g_{35} &= 1-u_3,& g_{45} &= 1,&
\nonumber\\[1ex]
g_{16} &= u_6,& g_{26} &= 1,&
g_{36} &= 1,& g_{46} &= 1-u_3,&
\nonumber\\[1ex]
g_{17} &= 1,& g_{27} &= 1-u_7,&
g_{37} &= 1,& g_{47} &= 1,&
\nonumber\\[1ex]
\end{align}
and
\vspace{5pt}
\begin{align}
g_{51} &= \frac{(1-u_4)^2}{u_3 u_4 u_5 u_6},&
g_{61} &= \frac{u_6 (1-u_3)^2}{u_1 u_2 u_3 (1-u_4)},&
\nonumber\\[1ex]
g_{52} &= \frac{u_4 u_5 u_6}{(1-u_4)^2},&
g_{62} &= \frac{u_1 u_2 u_3 u_4}{(1-u_3)^2},&
\nonumber\\[1ex]
g_{53} &= \frac{u_1 (1-u_4)^2}{u_5 u_6 (1-u_3)},&
g_{63} &= \frac{(1-u_3)^2(1-u_4)}{u_1 u_2 u_3 u_4 u_5},&
\nonumber\\[1ex]
g_{54} &= \frac{u_6 (1-u_3)^2}{u_1 u_2 (1-u_4)},&
g_{64} &= \frac{u_2 u_3 u_4 u_5 u_6}{(1-u_3)(1-u_4)^2},&
\nonumber\\[1ex]
g_{55} &= \frac{u_1 u_2 u_3}{(1-u_3)^2},&
g_{65} &= \frac{(1-u_4)^2}{u_3 u_4 u_5 u_6},&
\nonumber\\[1ex]
g_{56} &= \frac{(1-u_3)^2}{u_1 u_2 u_3 u_4},&
g_{66} &= \frac{u_4 u_5 u_6 (1-u_3)}{u_1 (1-u_4)^2},&
\nonumber\\[1ex]
g_{57} &= \frac{u_2 u_3 u_4 u_5 (1-u_7)}{(1-u_3)^2(1-u_4)^2},&
g_{67} &= \frac{u_1 u_2(1-u_4)^2}{u_5 u_6 (1-u_3)^2}.
\nonumber\\[1ex]
\end{align}
These formulae give the following reduced symbol alphabet on \CO{3}{4},
\be
\label{eq:COletters}
\{u_1,\,u_2,\,u_3,\,1-u_3,\,u_4,\,1-u_4,\,u_5,\,u_6,\,1-u_7 \} \,.
\ee
Alphabets on the other CO surfaces can be obtained by dihedral plus parity transformations.

\subsection{Infinitesimal line}
\label{app:IL}

To obtain a twistor parametrization for the IL~(\ref{eq:IL}), we employ of the soft limit parametrization~\cite{DelDuca:2016lad,Dixon:2020cnr}
\be\label{eq:ZsSoft}
Z_1 = c_6 Z_6 + c_7 Z_7 + c_2 Z_2 + c_3 Z_3 \,,
\ee
with $c_3,c_6\to0$ and $Z_2\ldots Z_7$ the six-point momentum twistors, and then take the soft limit onto the six-point origin kinematics \cite{Basso:2020xts}. In this limit, the $g$-letters collapse to the following,
\vspace{5pt}
\begin{align}
\label{eq:gsIL}
g_{11} &= u_1,& g_{21} &= 1,&
g_{31} &= 1,& g_{41} &= 1-u_4 u_7,&
\nonumber\\[1ex]
g_{12} &= u_2,& g_{22} &= 1,&
g_{32} &= 1-u_4 u_7,& g_{42} &= 1,&
\nonumber\\[1ex]
g_{13} &= u_3,& g_{23} &= 1,&
g_{33} &= 1,& g_{43} &=1-u_4 u_7,&
\nonumber\\[1ex]
g_{14} &= 1,& g_{24} &= 1-u_4,&
g_{34} &= 1,& g_{44} &= 1,&
\nonumber\\[1ex]
g_{15} &= u_5,& g_{25} &= 1,&
g_{35} &= 1,& g_{45} &= 1,&
\nonumber\\[1ex]
g_{16} &= u_6,& g_{26} &= 1,&
g_{36} &= 1,& g_{46} &= 1,&
\nonumber\\[1ex]
g_{17} &= 1,& g_{27} &= 1-u_7,&
g_{37} &= 1,& g_{47} &= 1,&
\nonumber\\[1ex]
\end{align}
and
\vspace{5pt}
\begin{align}
\label{eq:gsIL2}
g_{51} &= \frac{(1-u_4)^2 u_6}{(1-u_7)^2 u_3 u_5},&
g_{61} &= \frac{(1-u_7)^2}{(1-u_4 u_7) u_1 u_2 u_3 u_6},&
\nonumber\\[1ex]
g_{52} &= \frac{(1-u_7)^2 u_5}{(1-u_4)^2 u_6},&
g_{62} &= u_1 u_2 u_3 ,&
\nonumber\\[1ex]
g_{53} &= \frac{(1-u_4)^2 u_1 u_6}{(1-u_7)^2 u_5},&
g_{63} &= \frac{(1-u_4)^2}{(1-u_4 u_7) u_1 u_2 u_3 u_5},&
\nonumber\\[1ex]
g_{54} &= \frac{(1-u_7)^2}{(1-u_4) u_1 u_2 u_6},&
g_{64} &= \frac{(1-u_7)^2 u_2 u_3 u_5}{(1-u_4)^2 u_6},&
\nonumber\\[1ex]
g_{55} &= u_1 u_2 u_3 ,&
g_{65} &= \frac{(1-u_4)^2 u_6}{(1-u_7)^2 u_3 u_5},&
\nonumber\\[1ex]
g_{56} &= \frac{1}{u_1 u_2 u_3},&
g_{66} &= \frac{(1-u_7)^2 u_5}{(1-u_4)^2 u_1 u_6},&
\nonumber\\[1ex]
g_{57} &= \frac{(1-u_7) u_2 u_3 u_5}{(1-u_4)^2},&
g_{67} &= \frac{(1-u_4)^2 u_1 u_2 u_6}{(1-u_7)^2 u_5}.
\nonumber\\[1ex]
\end{align}
with the reduced symbol alphabet
\be
\label{eq:ILletters}
\{ u_1,\,u_2,\,u_3,\,1-u_4,\,u_5,\,u_6,\,1-u_7,\,1-u_4 u_7 \} \,.
\ee
This limit is a special case of the general soft limit discussed in Appendix D
of Ref.~\cite{Dixon:2020cnr}. However, the letters $1-u_4,1-u_7,$ and $1-u_4 u_7$ do not drop out of the heptagon functions here, because we are on the physical Riemann sheet, whereas the analogous letters do drop out on the Euclidean sheet~\cite{Dixon:2020cnr}.

\subsection{MRK}

We can obtain a parametrization for MRK with the scaling
\be
T_j \mapsto T_j \cdot \epsilon,  \qquad
S_j \mapsto \frac{S_j}{\epsilon}
\ee
in the OPE parametrization~(\ref{eq:ZsOPE}). In the limit $\epsilon\to0$,
the OPE variables are related to $\tau_1,\tau_2,z_1,z_2$ by
\bea
\tau_1 &=& \frac{T_2}{S_2} \,, \qquad \tau_2\ =\ \frac{T_1}{S_1} \,,
\label{eq:tautoTSF}\\
z_1 &=& -\frac{S_2 T_2}{F_2} \,,
\qquad \bar{z}_1\ =\ -S_2 T_2 F_2 \,,
\qquad z_2\ =\ -\frac{F_1}{S_1 T_1} \,,
\qquad \bar{z}_2\ =\ -\frac{1}{S_1 T_1 F_1} \,.
\label{eq:ztoTSF}
\eea
Taking the above limit and changing variables to the simplicial coordinates~(\ref{eq:simplicial}), the $g$-letters become the following,
\vspace{5pt}
\begin{align}
\label{eq:gsMRK}
g_{11} &= \tau_1 \left|\frac{\rho_1-\rho_2}{\rho_1(1-\rho_2)}\right|,&
g_{21} &= 1,&
g_{31} &= 1,&
\nonumber\\[1ex]
g_{12} &= \tau_1\left|\frac{\rho_1(1-\rho_2)}{\rho_1-\rho_2}\right|,&
g_{22} &= 1,&
g_{32} &= \tau_2 \frac{|1-\rho_2|^2}{|(1-\rho_1)(\rho_1-\rho_2)|},&
\nonumber\\[1ex]
g_{13} &= 1,&
g_{23} &= \tau_1 \frac{|\rho_2(1-\rho_1)|^2}{|\rho_1(1-\rho_2)(\rho_1-\rho_2)|},&
g_{33} &= 1,&
\nonumber\\[1ex]
g_{14} &= 1,&
g_{24} &= \tau_2 \frac{|1-\rho_2|^2}{|(1-\rho_1)(\rho_1-\rho_2)|},&
g_{34} &= 1,&
\nonumber\\[1ex]
g_{15} &= \tau_2 \left|\frac{1-\rho_1}{\rho_1-\rho_2}\right|,&
g_{25} &= 1,&
g_{35} &= \tau_1 \frac{|\rho_2(1-\rho_1)|^2}{|\rho_1(1-\rho_2)(\rho_1-\rho_2)|},&
\nonumber\\[1ex]
g_{16} &= \tau_2\left|\frac{\rho_1-\rho_2}{1-\rho_1}\right|,&
g_{26} &= 1,&
g_{36} &= 1,&
\nonumber\\[1ex]
g_{17} &= 1,&
g_{27} &= \tau_1 \tau_2 \left|\frac{(1-\rho_1)(1-\rho_2)}{\rho_1}\right|,&
g_{37} &= 1,&
\nonumber\\[1ex]
\end{align}
%
and
\vspace{5pt}
\begin{align}
\label{eq:gsMRK2}
g_{41} &= \tau_2 \frac{|1-\rho_2|^2}{|(1-\rho_1)(\rho_1-\rho_2)|},&
g_{51} &= \frac{(1-\bar{\rho}_1)(\rho_1-\rho_2)}{(1-\rho_1)(\bar{\rho}_1-\bar{\rho}_2)},&
g_{61} &= \frac{\bar{\rho}_1}{\rho_1},&
\nonumber\\[1ex]
g_{42} &= 1,&
g_{52} &= \frac{(1-\rho_1)(\bar{\rho}_1-\bar{\rho}_2)}{(1-\bar{\rho}_1)(\rho_1-\rho_2)},&
g_{62} &= \frac{\rho_1(1-\rho_2)(\bar{\rho}_1-\bar{\rho}_2)}{\bar{\rho}_1(1-\bar{\rho}_2)(\rho_1-\rho_2)} ,&
\nonumber\\[1ex]
g_{43} &= \tau_2 \frac{|1-\rho_2|^2}{|(1-\rho_1)(\rho_1-\rho_2)|},&
g_{53} &= \frac{\rho_2}{\bar{\rho}_2},&
g_{63} &= \frac{\bar{\rho}_1(1-\bar{\rho}_1)(\rho_1-\rho_2)}{\rho_1(1-\rho_1)(\bar{\rho}_1-\bar{\rho}_2)},&
\nonumber\\[1ex]
g_{44} &= \tau_1 \frac{|\rho_2(1-\rho_1)|^2}{|\rho_1(1-\rho_2)(\rho_1-\rho_2)|},&
g_{54} &= \frac{\bar{\rho}_1}{\rho_1},&
g_{64} &= \frac{\rho_1 \bar{\rho}_2 (1-\rho_2)(\bar{\rho}_1-\bar{\rho}_2)}{\bar{\rho}_1 \rho_2 (1-\bar{\rho}_2)(\rho_1-\rho_2)},&
\nonumber\\[1ex]
g_{45} &= 1,&
g_{55} &= \frac{\rho_1(1-\rho_2)(\bar{\rho}_1-\bar{\rho}_2)}{\bar{\rho}_1(1-\bar{\rho}_2)(\rho_1-\rho_2)} ,&
g_{65} &= \frac{(1-\bar{\rho}_1)(\rho_1-\rho_2)}{(1-\rho_1)(\bar{\rho}_1-\bar{\rho}_2)},&
\nonumber\\[1ex]
g_{46} &= \tau_1 \frac{|\rho_2(1-\rho_1)|^2}{|\rho_1(1-\rho_2)(\rho_1-\rho_2)|},&
g_{56} &= \frac{\bar{\rho}_1(1-\bar{\rho}_2)(\rho_1-\rho_2)}{\rho_1(1-\rho_2)(\bar{\rho}_1-\bar{\rho}_2)},&
g_{66} &= \frac{\bar{\rho}_2}{\rho_2},&
\nonumber\\[1ex]
g_{47} &= 1,&
g_{57} &= \frac{\rho_1(\bar{\rho}_1-\bar{\rho}_2)}{\bar{\rho}_1(\rho_1-\rho_2)},&
g_{67} &= \frac{\rho_1(1-\rho_2)(1-\bar{\rho}_1)}{\bar{\rho}_1(1-\bar{\rho}_2)(1-\rho_1)} .
\nonumber\\[1ex]
\end{align}
%
From the above relations, we see that the symbol alphabet in MRK is
\be \label{eq:MRKletters}
\{ \tau_1,\, \tau_2,\,
\rho_1,\, \bar{\rho}_1,\, 1-\rho_1,\, 1-\bar{\rho}_1,\,
\rho_2,\, \bar{\rho}_2,\, 1-\rho_2,\, 1-\bar{\rho}_2,\,
\rho_1-\rho_2,\, \bar{\rho}_1-\bar{\rho}_2 
\}\,.
\ee


\bibliographystyle{JHEP}
\bibliography{heptagon}

\end{document}